\documentclass[%
 reprint,
 superscriptaddress,
 amsmath,amssymb,
 prx,longbibliography
]{revtex4-1}
\usepackage{graphicx}% Include figure files
\usepackage{dcolumn}% Align table columns on decimal point
\usepackage{bm}% bold math
\usepackage{float}
\usepackage{siunitx}
\usepackage{color}
\usepackage{amsmath}
\usepackage{mathtools}
\usepackage{amssymb}
\usepackage[colorlinks]{hyperref}% add hypertext capabilities
\hypersetup{%
    plainpages=true,
    breaklinks=true,% not default in dvips mode, so we must specify
    hypertexnames=false,%not ideal, but needed when pagenums duplicate (`i' vs. `1')
    pageanchor=true,
    colorlinks=true,
    linkcolor={black},
    citecolor={blue},
    urlcolor={blue},
    anchorcolor={black}
}

\usepackage{array}
\newcolumntype{L}[1]{>{\raggedright\let\newline\\\arraybackslash\hspace{0pt}}p{#1}}
\newcolumntype{C}[1]{>{\centering\let\newline\\\arraybackslash\hspace{0pt}}p{#1}}
\newcolumntype{R}[1]{>{\raggedleft\let\newline\\\arraybackslash\hspace{0pt}}p{#1}}

\usepackage[utf8]{inputenc}
\usepackage{dsfont}
\usepackage{cleveref}
\usepackage{xcolor}
\usepackage{pgf}
\usepackage{placeins}

\usepackage{newfloat}

\AtBeginDocument{%
    \newwrite\bibnotes
    \def\bibnotesext{Notes.bib}
    \immediate\openout\bibnotes=\jobname\bibnotesext
    \immediate\write\bibnotes{@CONTROL{REVTEX41Control}}
    \immediate\write\bibnotes{@CONTROL{%
    apsrev41Control,author="08",editor="1",pages="1",title="0",year="1"}}
     \if@filesw
     \immediate\write\@auxout{\string\citation{apsrev41Control}}%
    \fi
}%
\begin{document}

\preprint{APS/123-QED}

\title{Deterministic remote entanglement using a chiral quantum interconnect}

%TITLE TRIALS
% Deterministic Remote Entanglement using Tunably Chiral Quantum Electrodynamics
% Deterministic remote entanglement using chiral microwave photons in an all-to-all-compatible architecture
% Deterministic remote entanglement through a chiral quantum interconnect in an all-to-all-compatible architecture
% Deterministic entanglement of remote superconducting qubit modules in an all-to-all-compatible architecture
% Deterministic entanglement between superconducting circuits connected with a chiral waveguide
% Deterministic remote entanglement between superconducting modules in a scalable waveguide quantum electrodynamics platform
% Demonstration of a chiral quantum interconnect in an all-to-all-compatible architecture

\def\RLEaffil{Research Laboratory of Electronics, Massachusetts Institute of Technology, Cambridge, MA 02139, USA}
\def\LLaffil{Lincoln Laboratory, Massachusetts Institute of Technology, Lexington, MA 02421, USA}
\def\Physaffil{Department of Physics, Massachusetts Institute of Technology, Cambridge, MA 02139, USA}
\def\EECSaffil{Department of Electrical Engineering and Computer Science, Massachusetts Institute of Technology, Cambridge, MA 02139, USA}
\def\affilAQ{\textit{Atlantic Quantum, Cambridge, MA 02139}}
\def\affilGoogle{\textit{Google Quantum AI, Goleta, CA 93111}}

\author{Aziza~Almanakly}
\email{azizaalm@mit.edu}

% \author{Agustin Di Paolo}
% \affiliation{\RLEaffil}

\author{Beatriz~Yankelevich}
\affiliation{\RLEaffil}
\affiliation{\EECSaffil}

\author{Max~Hays}
\affiliation{\RLEaffil}

\author{Bharath~Kannan}
\altaffiliation[Present address: ]{\affilAQ}
\affiliation{\RLEaffil}
\affiliation{\EECSaffil}

\author{R\'eouven~Assouly}
\affiliation{\RLEaffil}

\author{Alex~Greene}
\altaffiliation[Present address: ]{\affilGoogle}
\affiliation{\RLEaffil}
\affiliation{\EECSaffil}

\author{Michael~Gingras}
\author{Bethany~M.~Niedzielski}
\author{Hannah~Stickler}
\affiliation{\LLaffil}

\author{Mollie~E.~Schwartz}
\affiliation{\LLaffil}

\author{Kyle~Serniak}
\affiliation{\RLEaffil}
\affiliation{\LLaffil}

\author{Joel~ I-J.~Wang}
\affiliation{\RLEaffil}

\author{Terry~P.~Orlando}
\affiliation{\RLEaffil}
\affiliation{\EECSaffil}

\author{Simon~Gustavsson}
\altaffiliation[Present address: ]{\affilAQ}
\affiliation{\RLEaffil}

\author{Jeffrey~A.~Grover}
\affiliation{\RLEaffil}
 
\author{William~D.~Oliver}
\email{william.oliver@mit.edu}
\affiliation{\RLEaffil}
\affiliation{\EECSaffil}
\affiliation{\Physaffil}

\begin{abstract}
Quantum interconnects facilitate entanglement distribution between non-local computational nodes.
For superconducting processors, microwave photons are a natural means to mediate this distribution.
However, many existing architectures limit node connectivity and directionality.
In this work, we construct a chiral quantum interconnect between two nominally identical modules in separate microwave packages.
We leverage quantum interference to emit and absorb microwave photons on demand and in a chosen direction between these modules.
We optimize the protocol using model-free reinforcement learning to maximize absorption efficiency.
By halting the emission process halfway through its duration, we generate remote entanglement between modules in the form of a four-qubit $W$ state with $62.4 \pm 1.6 \%$ (leftward photon propagation) and $62.1 \pm 1.2\%$ (rightward) fidelity, limited mainly by propagation loss.
A chiral quantum network comprising many modules provides a platform for the exploration of novel many-body physics and quantum simulation.
This quantum network architecture enables all-to-all connectivity between non-local processors for modular and extensible quantum~computation.
\end{abstract}
\maketitle
\section{Introduction}
Quantum computation will likely rely on networks that distribute entanglement throughout the computer as system size scales~\cite{Cirac1997, Cirac1999, Kimble2008}. 
Such a network comprises many quantum interconnects, the channels by which quantum information is transmitted between non-local processing nodes.
Examples of experimentally realized quantum interconnects include optical-photon propagation through fibers between trapped ions~\cite{Monroe2014, Hucul2015} or spin qubits in diamond vacancy centers~\cite{Hermans2022,Hensen2015,stolk2024,Knaut2024}, coupling distant semiconductor spin qubits with superconducting circuit elements~\cite{dijkema2023, Pita_Vidal_2024}, and  neutral-atom shuttling across 2D atomic arrays~\cite{bluvstein2021quantum}.
Additionally, non-local qubit connectivity would facilitate the implementation of high-rate low-density parity check codes with lower resource requirements associated with their non-local error detection and correction protocols~\cite{Bravyi2024, Breuckmann2021}. 

Within superconducting processors, microwave photons are a natural means to transmit quantum information between modules.
In one form of demonstrated communication, bound states in a microwave resonator couple distant qubits~\cite{Zhong2019,Leung2019,Chang2020,Zhong2021,Burkhart2021}.
State-transfer protocols using this type of interconnect have been demonstrated with high fidelity~\cite{qiu2023, Niu2023}, but are constrained by the length of the resonator.
Interconnects that instead use itinerant microwave photons in waveguides are not constrained in length, but often require circulators that enforce unidirectional communication and introduce loss~\cite{Kurpiers2018, Axline2018, Campagne-Ibarcq2018, Kurpiers2019, Magnard2020}.
Both schemes are restricted in connectivity, requiring successive transfers with compounding error rates and constraining future quantum computer architectures as the number of modules increases.

Waveguide quantum-electrodynamical (wQED) systems~\cite{Lalumiere2013, Sheremet2023}, in which atoms strongly couple to a photonic continuum in a 1D waveguide, are a promising platform for realizing extensible quantum networks~\cite{Solano2017_superradiance, Gheeraert2020, Guimond2020, Solano2023}. 
Due to the large electric dipole moment of superconducting qubits, the strong-coupling regime of wQED---where the qubit-waveguide coupling is higher than intrinsic qubit losses---is straightforwardly accessible with coupling efficiencies higher than 99$\%$  ~\cite{Wallraff2004}.
As a consequence, phenomena such as resonance fluorescence~\cite{Astafiev2010,Hoi2011,Hoi2013,Hoi2015}, super- and sub-radiance~\cite{Dicke1954,VanLoo2013,Mirhosseini2019,Zanner2022}, and deterministic microwave single-photon emission~\cite{Abdumalikov2011,Hoi2012,Forn-Diaz2017,Gonzalez-Tudela2015,Pfaff2017,Gasparinetti2017,besse2020realizing,Kannan2020, Ferreira2024} have been demonstrated. 

Chiral wQED focuses on interactions between atoms and a 1D photonic continuum that are direction-dependent in nature~\cite{Pichler2015, Lodahl2017}.
A qubit is chirally coupled to a 1D waveguide if it interacts more strongly with either the leftward- or rightward-propagating photonic modes.
Chiral coupling of an atom to an optical nanofiber has been demonstrated using spin-polarization locking~\cite{Bliokh2015,Petersen2014,Mitsch2014,Sollner2015,Coles2016}.
Consequences of chirality include directional single-photon emission and driven-dissipative entanglement schemes independent of distance~\cite{Stannigel2012,Pichler2015,Soro2022,Joshi2023,lingenfelter2024,irfan2024}.

\begin{figure*}[t!]
    \centering
    \includegraphics[width=\textwidth]{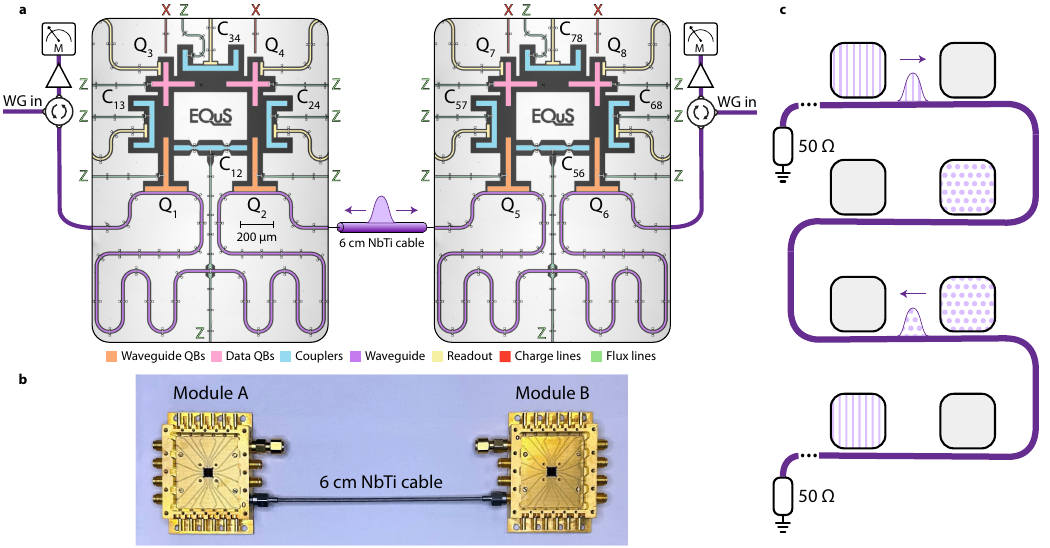}
    \caption{
    \textbf{Chiral quantum interconnect experimental set-up.}
    \textbf{a)} False-colored optical micrographs of the devices and experimental schematic.
    Each module is capable of directional photon emission and absorption. The tunable couplers (blue) mediate interactions between pairs of data qubits (pink) and waveguide qubits (orange), which serve to emit and absorb directional photons to and from the waveguide. The photons propagate through a waveguide (purple) terminating in a measurement chain at both ends. Waveguide in (WG in) denotes microwave ports that enable an independent assessment of losses due to wirebonds, connectors, and normal-metal traces (see Supplementary Information). X and Z correspond respectively to qubit transverse and longitudinal drives.
    \textbf{b)} Representative photograph of experimental implementation of the interconnect.
    Each module is housed in a separate microwave package.
    The two packages are connected with a 6 cm NbTi superconducting cable through which directional photons propagate.
    This composite system is then operated at millikelvin temperatures in a dilution refrigerator utilizing standard techniques.
    \textbf{c)} Conceptual diagram of an all-to-all quantum network using chiral wQED.
    Modules are tiled along an arbitrarily long, one-dimensional waveguide and in principle serve as interfaces to individual quantum processing nodes.
    Directional photons mediate communication between modules, which function in three modes: emission, absorption, and transparency by adjusting the frequencies of the waveguide qubits (on-resonance with the emitted photon for absorption, and off-resonance for transparency).
    We show an example of two communication protocols, where the dotted and striped pairs of modules each communicate via microwave photons propagating in opposite directions, bypassing transparent (gray) modules in between.
    With this scheme, we can distribute entanglement between modules on the waveguide.
    }
    \label{fig:fig1}
\end{figure*}
Recently in superconducting systems, deterministic directional emission has been realized using quantum interference; the simultaneous emission of two entangled qubits a specific distance apart along a bidirectional waveguide results in photon propagation in a chosen direction~\cite{kannan2023,Gheeraert2020}.
The time-reversed process absorbs directional photons via the same interference effect.
Together, directional emission and absorption form a new method of quantum communication across essentially any arbitrary distance, limited only by loss along the waveguide.
Tunable chiral qubit-waveguide coupling via quantum interference is the physical basis of our interconnect.

In this experiment, we construct an interconnect between two superconducting modules that communicate via directional emission and absorption of photons to and from a common open waveguide, as shown in Fig.~\ref{fig:fig1}a.
Each module is housed in its own microwave package, as photographed in Fig.~\ref{fig:fig1}b.
At times and in directions of our choosing, photons propagate back and forth between the modules through a 6 cm superconducting cable.
In principle, this waveguide, with 50 $\Omega$ terminations on both ends, can extend an arbitrary length and host simultaneous communication between any pair of~modules. 

To optimize the directional photon emission and absorption protocol, we implement a model-free reinforcement learning (RL) pulse optimization algorithm~\cite{sivak2022, sivak2023, ding2023}.
First, we characterize the protocol in both propagation directions by measuring the excited state populations of qubits on the absorber modules.
Then we measure the field amplitudes of the itinerant directional photons with heterodyne detection at both ends of the waveguide~\cite{Eichler2011,Eichler2012,Lang2013,kannanN00N2020}. 
Finally, we perform quantum state tomography to characterize the resulting qubit states following the protocol.
By halting emission halfway and executing the RL-optimized absorption protocol, we create remotely~entangled~four-qubit $W$ states with $62.4 \pm 1.6 \%$ (leftward photon propagation) and $62.1 \pm 1.2\%$ (rightward) fidelity~\cite{Dur2000}, well above the threshold necessary for purification protocols yielding higher fidelity entangled states~\cite{Bennet1996,Yan2022, Gidney2023, ramette2023}.

\subsection{Chiral Emission and Absorption Protocol}
Each module contains two frequency-tunable transmon waveguide qubits (Q$_{1/2}$ and Q$_{5/6}$)~\cite{Koch2007}, which are directly coupled to a common coplanar waveguide at rate $\gamma/2\pi \approx 17$ MHz.
We separate the qubits along the waveguide by a distance of $\Delta x = \lambda/4$, where $\lambda = 2\pi/k$ is the wavelength of the emission at the qubit frequency $\omega_{1,2,5,6}/2\pi = 5.0$ GHz -- all four qubits on both modules must be resonant.
The two modules are roughly $d\approx$ 10 cm apart along the waveguide, but in principle they can be separated by any distance.

% Because the inter-module distance $d \approx 10\ \mathrm{cm} \ll \nu/\gamma$, where $\nu$ is the speed of light in the waveguide, we can model our system with the master-equation formalism in the Markov limit (See Supplementary Info). In this limit, two emitter qubits interact with each other through the waveguide modes in one of two ways: waveguide-mediated exchange with strength $ \frac{\gamma}{2} \sin k(x_j - x_i)$, or correlated dissipation with strength $ \gamma\cos k(x_j - x_i)$, where $x_{i/j}$ references the position of each qubit along the waveguide~\cite{Lalumiere2013}.
% Waveguide-mediated exchange serves as a coherent coupling between the emitter qubits, while correlated dissipation signifies interference in spontaneous emission.
% Precise control of these interactions is critical to achieving chiral photon emission and absorption. 
To emit a directional photon from one module to another, we exploit the interference of the two-waveguide-qubit state as it decays into the waveguide.
The input-output equations for this system of two uncoupled waveguide qubits Q$_1$ and Q$_2$ elucidate the quantum interference effect that results in directional emission and absorption~\cite{Lalumiere2013,Gheeraert2020},
\begin{equation}
\label{eq:inout}
\begin{split}
        &\hat{a}_\textrm{L}^{} = \hat{a}_\textrm{L}^\textrm{in} + \sqrt{\frac{\gamma}{2}} (\hat{\sigma}_1^- + \hat{\sigma}_2^-e^{ik\Delta x}), \\
        &\hat{a}_\textrm{R}^{} = \hat{a}_\textrm{R}^\textrm{in} + \sqrt{\frac{\gamma}{2}} (\hat{\sigma}_1^- + \hat{\sigma}_2^-e^{-ik\Delta x}),
\end{split}
\end{equation}
where $\hat{a}_\textrm{L(R)}^\textrm{in}$ represents an input field in the leftward (rightward) propagating mode, and $\hat\sigma_{1/2}^-$ are the waveguide-qubit lowering operators.
These equations relate the initial state of the waveguide qubits to the average output photon flux $\langle\hat{n}_\textrm{L(R)}\rangle = \langle \hat{a}_\textrm{L(R)}^\dagger\hat{a}_\textrm{L(R)}\rangle$ in both propagation directions.
In the case of zero input field, if the waveguide qubits are initialized in the entangled state $|\psi^+\rangle = (|eg\rangle + e^{ik\Delta x} |ge\rangle)/\sqrt{2}$, the resulting photon fluxes at both ends of the waveguide are $\langle\hat{n}_\textrm{R}\rangle =\gamma$ and $\langle\hat{n}_\textrm{L}\rangle = 0$; the decay of the two-qubit entangled state $|\psi^+\rangle$ thus results in a rightward-propagating photon. Similarly, the initialization of the qubits into  $|\psi^-\rangle = (|eg\rangle + e^{-ik\Delta x} |ge\rangle)/\sqrt{2}$ results in a leftward propagating photon.
The input-output equations are also a tool to describe directional photon absorption; the creation of a photon in the leftward or rightward mode results in the corresponding entangled state $\hat a_\mathrm{R/L}^\dagger|gg\rangle \to |\psi ^\pm\rangle$. Any interaction between the waveguide qubits disrupts the interference effect; therefore, we use a transmon tunable coupler (C$_{12}$ and C$_{56}$)~\cite{yan2018,sung2020} to cancel the waveguide-mediated exchange coupling~\cite{Lalumiere2013}.

% Resonant qubits sharing a waveguide exhibit two types of distance-dependent interactions: waveguide-mediated coherent coupling of the form $\gamma/2\sin \left(k\Delta x\right) [\hat \sigma_1 ^+ \hat \sigma_2^- + \hat \sigma_1 ^- \hat \sigma_2^+]$, and correlated dissipation of the form $-\gamma \cos \left(k\Delta x\right) D[\hat \sigma_1^-, \hat \sigma_2^-]$, where $D[\hat A, \hat B] = \hat{B}\hat\rho\hat{A}^\dagger - \frac{1}{2} \{\hat{B}^\dagger\hat{A},\hat\rho\}$ is the Lindblad dissipator~\cite{Lalumiere2013}.
% At a separation of $\Delta x = \lambda/4$, the strength of correlated dissipation is zero, while the emitter qubits have a maximal coherent coupling of $\gamma/2$, which we cancel with a transmon tunable coupler (C$_{12}$ and C$_{56}$)~\cite{yan2018,sung2020}.
% Under these calibration conditions, the emitter qubits are uncoupled as they decay into the waveguide.
\begin{figure}[t!]
    \centering
    \includegraphics[width=3.45in]{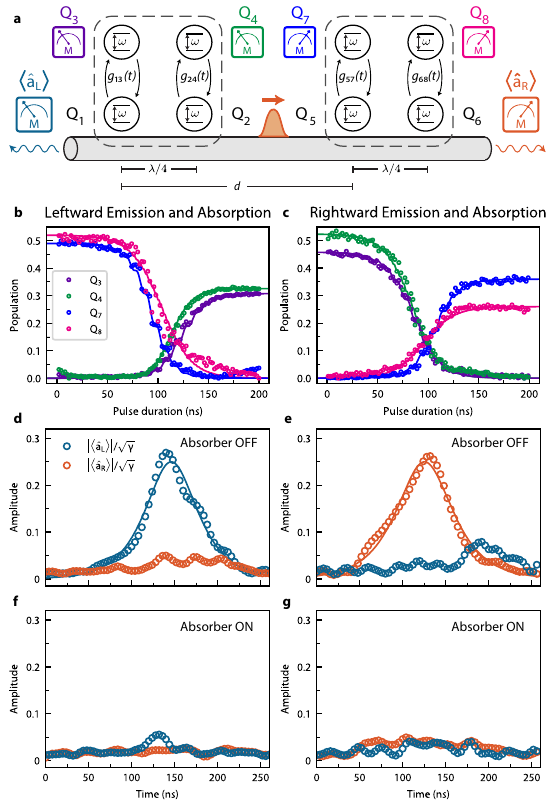}
    \caption{
    \textbf{Chiral photon emission and absorption in the time domain.}
    \textbf{a)} Schematic diagram of rightward photon emission experiment. We measure the population of all four data qubits and the photon field amplitude at both outputs of the waveguide for time-domain characterization. \textbf{b, c)} Data-qubit population as a function of the duration of the RL-shaped pulses that generate and capture a leftward- (rightward-) propagating photon. Leftward (rightward) absorption yields a maximum data qubit population of 64$\%$ (63$\%$). The solid lines represent fits to master equation simulations. \textbf{d, e)} Photon field amplitude measurements of emission in the waveguide. The data qubits on the emitter module are initialized in the state $(|gg\rangle + |\psi ^\pm\rangle)/\sqrt{2}$. After detuning the waveguide qubits on the absorber module from the photon frequency, we measure the emitted photon field amplitudes at both ends of the waveguide. The solid lines represent a fit to theory for a time-symmetric photon wavepacket of the form $\mathrm{sech}(\gamma_\mathrm{ph}t/2)$, where $\gamma_\mathrm{ph}/2\pi \approx $ 7 MHz is the photon linewidth. \textbf{f, g)} Photon field amplitudes in waveguide following the absorption experiment. The field amplitudes are diminished in both directions, illustrating absorption by the module.}
    \label{fig:fig2} 
\end{figure}
% The master equation that governs the dynamics of the four resonant emitter qubits on the waveguide is~\cite{Gheeraert2020, Lalumiere2013}
% \begin{equation}
%     \frac{d\hat{\rho}}{dt} = -i\big[\hat H_\mathrm{J} + \hat H_\mathrm{c}, \hat\rho\big] + \sum_{i,j} \gamma\cos(\phi_{i,j})D\big[\hat{\sigma}_i^-, \hat{\sigma}_j^-\big]\hat\rho,
%     \label{eq:ME_absorb}
% \end{equation}
% where $D[\hat{O}] = \hat{O}\hat\rho\hat{O}^\dagger - \frac{1}{2} \{\hat{O}^\dagger\hat{O},\hat\rho\}$ is the Lindblad dissipator.
% The waveguide-mediated exchange couplings between the four emitter qubits numbered $i, j \in \{1,2,5,6\}$ are described by $\hat{H_\mathrm{J}} =  \sum_{i,j} \frac{\gamma}{2}\sin(\phi_{i,j})[\hat{\sigma}_i^-\hat{\sigma}_j^+ +\hat{\sigma}_i^+\hat{\sigma}_j^-]$ The phase $\phi_{i,j} = k(x_j - x_i)$ represents the relative position phase between the emitter qubits along the waveguide.
% We induce static couplings $\hat{H}_c = -\frac{\gamma}{2} \hat \sigma^-_1 \hat \sigma^+_2 -\frac{\gamma}{2} \hat \sigma^-_5 \hat \sigma^+_6 + \mathrm{h.c.}  $ between emitter qubits on the same module with a transmon tunable coupler (C$_{12}$ and C$_{56}$)~\cite{yan2018,sung2020} in order to cancel intra-module waveguide-mediated exchange interactions.
% The emitter qubits on different modules also experience correlated dissipation dependent on the inter-qubit distance along the waveguide. 
Because the qubits are strongly coupled to the waveguide, we cannot prepare the desired entangled state with high fidelity solely using these qubits.
Instead, we introduce the data qubits (Q$_{3/4}$ and Q$_{7/8}$) and couple them to each other (C$_{34}$ and  C$_{78}$) and to the waveguide qubits (C$_{13}$, C$_{24}$, and C$_{57}$, C$_{68}$) with tunable couplers.
Because the data qubits are not subject to direct dissipation into the waveguide, we have time to prepare them into an entangled state with single- and two-qubit gates.
In an actual networking implementation, the data qubits would also serve as an interface to a local processor (not shown).

The complete chiral emission and absorption protocol begins by exciting one of the data qubits on the emitter module with a $\pi$-pulse.
We generate intra-module entanglement and perform photon swaps using parametric flux modulation of the tunable couplers (colored blue in Fig.~\ref{fig:fig1}a) at the frequency difference between their neighboring qubits.
To initialize the data qubits in the desired entangled state $|\psi^\pm\rangle = (|eg\rangle \pm i |ge\rangle)/\sqrt{2}$, we implement a $\sqrt{i\mathrm{SWAP}}$ gate. We realize the relative phases $\pm i$ by changing the phase of the coupler pulse modulation.

We transfer the entangled state from the data qubits to the waveguide qubits with a time-dependent coupling between each data-waveguide qubit pair, shaping the wavepacket of the emitted photon by adjusting the duration and envelope of each parametric drive.
In an ideal system, time-symmetric photon wavepackets simplify calibration because the absorption pulses are the time reverse of the emission pulses~\cite{Kurpiers2018, Yang2023}.  However, because of systemic nonidealities such as flux-control line distortions, pulse nonlinearities, AC Stark shifts, and impedance mismatches between and beyond modules~\cite{Roth2017, reuer2021realization, Yang2023, Rol2020}, the photon wavepacket experiences distortion.
We therefore seed a reinforcement learning agent with the ideal symmetric pulses and allow it the freedom to shape the pulses to maximize absorption efficiency~(see Supplementary Information).

Directional emission and absorption explicitly do not depend on the distance $d$ between modules -- a key feature of this interconnect in the context of extensible quantum networks. The utility of the chiral quantum interconnect is available at the local-chip scale for long-range coupling between qubits in addition to the distant-module scale demonstrated in this work.~In the language of chiral wQED, two modules form a cascaded system in which interactions are unidirectional and distance-independent (see Supplementary Information). Intuitively, a photon can propagate any arbitrary distance between modules, limited only by propagation loss along the waveguide.
\subsection{Qubit and Photon Measurements}
To characterize data-qubit population and photon-field amplitude in the waveguide during the emission and absorption processes, we perform a set of time-domain measurements.
We can detect the photon at each of the four data qubits or at either end of the waveguide throughout the protocol, as represented in Fig.~\ref{fig:fig2}a.
We simultaneously measure the population of all four data qubits as a function of the time duration of the emission and absorption pulses in Fig.~\ref{fig:fig2}b/c.
The data qubits on the emitter module start in the entangled state $|\psi^\pm\rangle$, and
as the duration of the pulses increases, the population is transferred to the data qubits on the absorber module.
We define absorption efficiency as the maximum total data-qubit population on the absorber module, which is roughly $63-64\%$ in both photon propagation directions.

At the same time, we visualize the photon throughout the protocol by measuring the average field amplitudes $\langle \hat a_\mathrm{L/R}(t) \rangle$ at either end of the waveguide.
We first detune the waveguide qubits of the absorber module so that they are transparent to the itinerant photon in the waveguide that subsequently arrives at either detector.
The field amplitudes average to zero due to the phase uncertainty of Fock states.
Therefore, we initially excite the data qubit on the emitter module with a $\pi/2$ pulse to produce emission that is an equal superposition of the photonic ground state and a single-photon Fock state. We detect the resulting phase-coherent emission with heterodyne voltage measurements.
\begin{figure}[t!]
    \centering
    \includegraphics[width=3.45in]{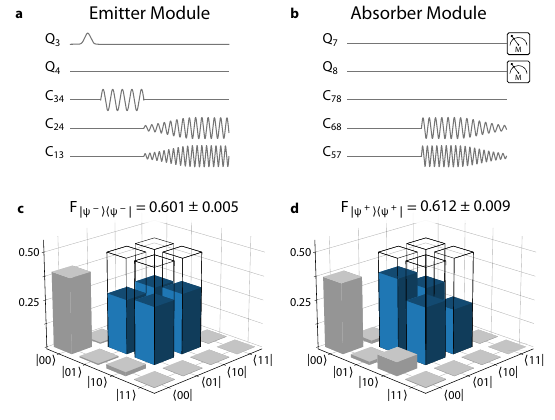}
    \caption{
    \textbf{Quantum state transfer between modules via directional photons.}
    \textbf{a)} Pulse sequence used to emit a rightward-propagating photon. An excitation is introduced on the emitter module with a $\pi$-pulse on the data qubit Q$_3$, followed by a parametric $\sqrt{\mathrm{iSWAP}}$ gate implemented with parametric flux modulation of the tunable coupler C$_{34}$ to prepare the data qubits into the entangled state $|\psi^+\rangle = (|eg\rangle + i |ge\rangle)/\sqrt{2}$. \textbf{b)} Pulse sequence for directional photon absorption. RL-optimized pulses are shaped to maximize absorption efficiency. At the end of the protocol, the data qubits on the absorber module should be in the same entangled state $|\psi^+\rangle$ initially prepared on the emitter module. \textbf{c, d)} Reconstructed density matrices of data qubit pairs after leftward (d, rightward) absorption. We measure Bell state fidelities $F_{|\psi ^ -\rangle \langle \psi ^ - |} = 0.601 \pm 0.005$ and $F_{|\psi ^ +\rangle \langle \psi ^ + |} = 0.612 \pm 0.009$ and concurrences of $C_\mathrm{L} = 0.597 \pm 0.005$ and $C_\mathrm{R} = 0.599 \pm 0.011$. We attribute absorption inefficiencies mainly to propagation loss and qubit decoherence. The magnitude of the ideal Bell state density matrix is shown with solid black lines.}
    \label{fig:fig3}
\end{figure}
\begin{figure*}[ht!]
    \centering
    \includegraphics[width=\textwidth]{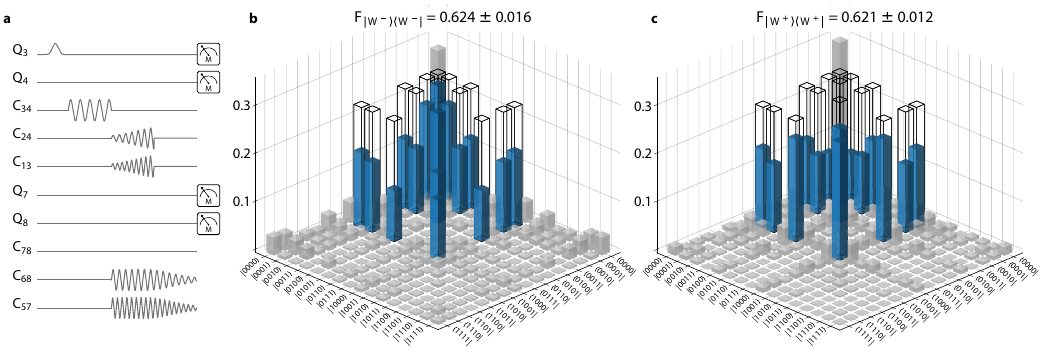}
    \caption{
    \textbf{Deterministic remote entanglement generation.}
    \textbf{a)} Pulse sequence to generate remote entanglement with rightward photon emission and absorption. We stop the RL-calibrated emission pulses on couplers C$_{13}$ and C$_{24}$ halfway through their duration, and implement the same full RL-calibrated absorption pulses used in Fig.~\ref{fig:fig3}. \textbf{b, c)} Reconstructed density matrices of all four data qubits after using leftward (rightward) photon emission and absorption to generate remote entanglement between distant modules. We measure $W$ state fidelities $F_{|W ^ -\rangle \langle W ^ - |} = 0.624 \pm 0.016$ and $F_{|W ^ +\rangle \langle W ^ + |} = 0.621 \pm 0.012$. The magnitude of the ideal state $|W^\pm\rangle = [|ggge\rangle \pm i |ggeg\rangle + e^{\mp ikd}(|gegg\rangle \pm i|eggg\rangle ) ]/2$ is shown with solid black lines, where $d$ is the distance between modules along the waveguide and $k = 2\pi/\lambda$.
    }
    \label{fig:fig4}
\end{figure*}

In Fig.~\ref{fig:fig2}d/e, we characterize the emitted photon field amplitudes $\langle \hat a_\mathrm{L/R}(t)\rangle$ as a result of the RL-optimized protocol pulses.
% The amplitude of the emitted photon is close to the theoretical prediction $\propto \mathrm{sech} (\gamma_\mathrm{ph}t/2)$, with deviations we attribute to the RL-optimized flux pulse envelopes.
% Manually, we can produce more symmetric photon shapes, but they do not result in the highest absorption efficiency. 
% The creation of time-symmetric photons is not a requirement for maximum absorption, but rather, it is the most straightforward calibration approach.
Presumably, the lack of time-symmetry of the photon amplitude measured at the detector counteracts experimental imperfections that diminish the absorption efficiency.
When absorbing the photon, we see the field amplitudes of the directional photon decrease significantly in both directions in Fig.~\ref{fig:fig2}f/g.
Notably, the field amplitude in the direction opposite communication also decreases following the absorption experiment, which could be indicative of reflections caused by impedance mismatches downstream of the modules, such as wirebonds, connectors, and circulators (see Supplementary Information).

At the end of the protocol, the data qubits on the absorber module should theoretically be in the same entangled state prepared on the emitter module $|\psi ^ \pm \rangle$. 
To measure the fidelity of the absorbed final entangled state, we implement two-qubit state tomography in both cases of leftward and rightward communication, resulting in the density matrices shown in Fig.~\ref{fig:fig3}c/d.
We measure Bell state fidelities of the data qubits on the absorber modules in both protocol directions $F_{|\psi ^ -\rangle \langle \psi ^ - |} = 0.601 \pm 0.005$ and $F_{|\psi ^ +\rangle \langle \psi ^ + |} = 0.612 \pm 0.009$ and concurrences of $C_\mathrm{L} = 0.597 \pm 0.005$ and $C_\mathrm{R} = 0.599 \pm 0.011$, quantifying the degree of entanglement. Error bars represent standard deviations of fidelities extracted from 100 repetitions of the tomography measurement, averaging each measurement over 50,000 shots. 

We categorize absorption inefficiencies into either coherent or incoherent photon loss. Using field-amplitude measurements, we characterize 4.2\% (7.2\%) coherent photon loss to the waveguide in the leftward (rightward) protocol directions. The main source of inefficiency comes in the form of 26.3\% incoherent photon loss predominantly due to propagation (scattering) loss between modules and data qubit~decoherence, which are not fundamental to the architecture (see Supplementary Information). By using a strictly superconducting waveguide and improving qubit design, we estimate that this protocol can achieve absorption efficiencies greater than 90\%.
At these projected efficiency levels, fault-tolerant modules can be connected without entanglement distillation~\cite{ramette2023}.

Addressing and minimizing incoherent losses will elucidate sources of coherent losses at the few percent level, which could be caused by impedance mismatches between and beyond the modules. By improving impedance matching with changes to the physical interconnect implementation and continuing reinforcement learning pulse optimization, we anticipate progression towards the demonstration of a high-fidelity directional emission and absorption protocol.

\subsection{Remote Entanglement Generation}
Once the absorption is optimized in both directions, we prepare entangled states spanning the two modules.
By halting the emission pulses halfway through their duration and executing full absorption pulses as illustrated in Fig.~\ref{fig:fig4}a, we should ideally produce a four-qubit $W$ state of the form $|W^\pm\rangle = [|ggge\rangle \pm i |ggeg\rangle + e^{\mp ikd}(|gegg\rangle \pm i|eggg\rangle ) ]/2$.
We carry out the experiment in both photon-propagation directions and perform four-qubit state tomography to characterize the density matrix and $W$ state fidelities $F_{|W ^ -\rangle \langle W ^ - |} = 0.624 \pm 0.016$ and $F_{|W ^ +\rangle \langle W ^ + |} = 0.621 \pm 0.012$.
Even in the presence of significant photon loss between modules, here primarily due to lossy materials in the packages, connectors, and the coaxial cable, our protocol produces remote multi-partite entanglement well above the fidelity threshold of $50\%$~\cite{Guhne2010}.
This protocol also exceeds the fidelity threshold for entanglement purification algorithms compatible with this experiment, which mitigate the effect of photon loss and produce higher fidelity entangled states~\cite{Bennet1996,Yan2022, Gidney2023, ramette2023}.
This approach can be extended further to entangle qubits on many modules with all-to-all connectivity.
\section{Conclusion}
We leverage tunably chiral waveguide quantum electrodynamics to create a quantum network architecture that supports all-to-all connectivity.
We construct a quantum interconnect within this architecture by connecting two modules that function as emitters and absorbers of directional photons propagating through a common waveguide.
We realize on-demand, chiral emission and absorption in both photon-propagation directions.
With model-free reinforcement learning, we optimize the pulse shapes and parameters for maximal absorption efficiency.
Finally, we generate remote entanglement between distant modules in the form of $W$ states, which enable communication schemes that are robust against photon~loss~\cite{Dur2001}.

The non-reciprocal interactions demonstrated in this work enable new explorations of driven-dissipative dynamics in light-matter interactions. High-fidelity, distance-independent remote entanglement can be achieved by continuously and simultaneously driving the modules through the waveguide. The extension of chiral driven-dissipation to a chain of many modules results in configurable multi-partite entanglement structures that are not accessible with bidirectional systems~\cite{Pichler2015}. Such chiral quantum networks enable exploration of many-body physics, photonic topological effects, new quantum phases of light and matter, and quantum simulation of complex systems\cite{Lodahl2017}.

The directional state transfer demonstrated with this interconnect has direct applications in quantum random-access memory architectures \cite{Weiss2024}.
The long-range coupling realized here is compatible with high-rate low-density parity check codes for resource-efficient error correction on superconducting quantum processors.
Natural next steps include the construction of low-loss interconnects~\cite{Niu2023} with bump-bonds ~\cite{Rosenberg2017, Yost2020} on multi-chip-modules~\cite{Field2024} to connect an essentially arbitrarily large number of modules with a single common waveguide.
This work enables the demonstration of all-to-all connectivity and many-module remote entanglement amenable to gate-teleportation schemes~\cite{Jiang2007,Chou2018} for large-scale distributed quantum computing. 
 
\section*{Acknowledgments}
This research was funded in part by the Army Research Office under Award No. W911NF-23-1-0045; in part by the AWS Center for Quantum Computing; and in part under Air Force Contract No. FA8702-15-D-0001.
AA acknowledges support from the P.D. Soros Fellowship program and the Clare Boothe Luce Graduate Fellowship.
BY acknowledges support from the Fannie and John Hertz Foundation and the NSF Graduate Research Fellowship Program.
MH is supported by an appointment to the Intelligence Community Postdoctoral Research Fellowship Program at MIT administered by Oak Ridge Institute for Science and Education (ORISE) through an interagency agreement between the U.S. Department of Energy and the Office of the Director of National Intelligence (ODNI).
 Any opinions, findings, conclusions or recommendations expressed in this material are those of the author(s) and do not necessarily reflect the views of the US Air Force or the US Government.

 \section*{Author Contributions}
AA designed the experiment procedure and conducted the measurements.
AA and BY designed the devices, performed theoretical calculations and simulations, analyzed data, and wrote the manuscript.
MH assisted in implementing reinforcement-learning optimization.
RA helped troubleshoot experiments and analyze data.
AG assisted with the automation of calibration.
MG, BMN, and HS fabricated the devices with coordination from KS and MES.
BY, BK, RA, and JI-JW assisted with the experimental setup.
TPO, SG, MH, JAG, and WDO supervised the project.
All authors discussed the results and commented on the manuscript.

\section*{Data Availability}
The data that support the findings of this study are available from the corresponding author upon reasonable request.

\section*{Code Availability}
The code used for numerical simulations and data analyses is available from the corresponding author upon reasonable request.

\bibliography{main}

\onecolumngrid
\newpage
\begin{center}
    \textbf{SUPPLEMENTARY INFORMATION}
\end{center}

\setcounter{figure}{0}
\setcounter{equation}{0}
\makeatletter 
\renewcommand{\thefigure}{S\@arabic\c@figure}
\renewcommand{\thetable}{S\@arabic\c@table}
\renewcommand{\theequation}{S\arabic{equation}}
\makeatother
\subsection{Device and Experimental Setup}

\newcolumntype{P}[1]{>{\centering\arraybackslash}p{#1}}

\begin{figure*}[h]
    \centering
    \includegraphics[width=\textwidth]{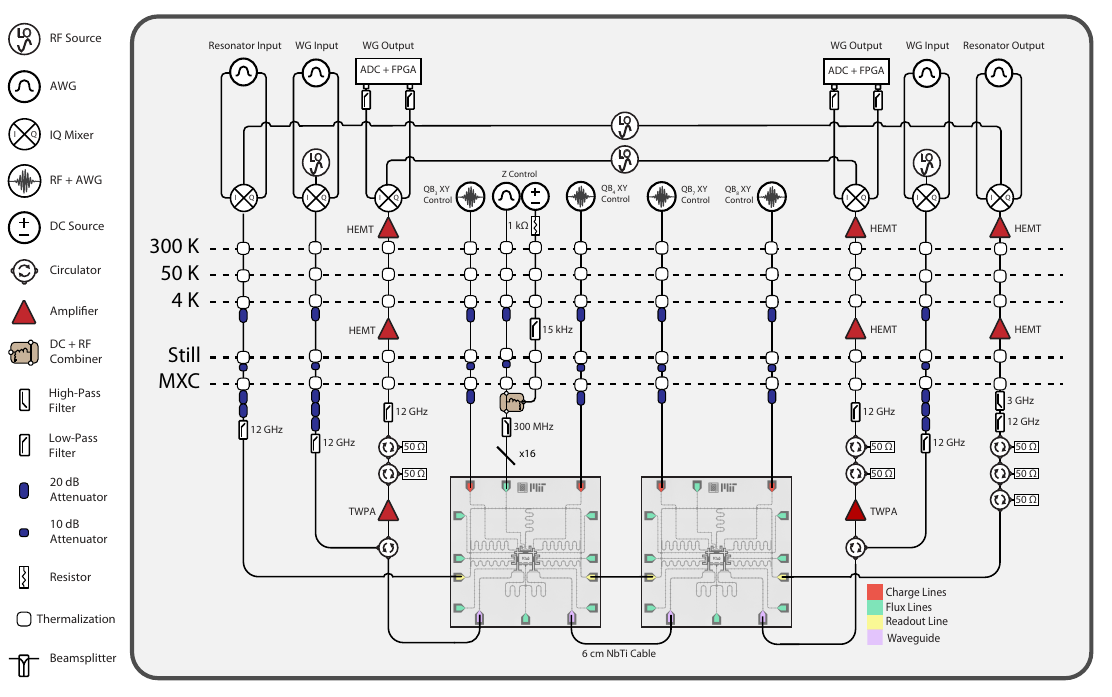}
    
    \caption{\textbf{Experimental setup.} Wiring schematic of the device and all electronics used to perform the experiment. Note that only one flux line configuration is shown (green), but each qubit and coupler is coupled to a flux line with separate, but identical, control electronics.}
    \label{fig:setup}
\end{figure*} 
This experiment was conducted in a Bluefors XLD1000 dilution refrigerator, which operates at base temperature of around 15 mK throughout the experiment.
The experimental setup is shown in Fig.~\ref{fig:setup}.
The device is protected from ambient magnetic fields by superconducting and Cryoperm-10 shields below the mixing chamber (MXC) stage.
Each end of the waveguide is connected to a microwave circulator for dual input-output operation.
To minimize thermal noise from higher temperature stages, the inputs are attenuated by 20 dB at the 4K stage, 10 dB at the Still stage, and 60 dB (40 dB for resonator readout input) at the MXC stage.
The output signals are filtered with $\SI{3}{GHz}$ high-pass and $\SI{12}{GHz}$ low-pass filters.
Two additional isolators are placed after the circulator at the MXC stage to prevent noise from higher-temperature stages travelling back into the sample. Traveling wave parametric amplifiers (TWPA) are used at the MXC stage and high electron mobility transistor (HEMT) amplifiers are used at $\SI{4}{K}$ and room-temperature stages of the measurement chain to amplify the outputs from the device.
The signals are then downconverted to an intermediate frequency using an IQ mixer, after which they are filtered, digitized, and demodulated.
All qubits and tunable couplers are also equipped with their own flux bias lines. A DC + RF combiner is used for all flux lines to provide both static and dynamic control of the qubit/coupler frequencies.
The DC and RF inputs are joined by a RF choke at the MXC stage before passing through a 300 MHz low-pass filter.
The RF flux control lines are attenuated by 20 dB at the $\SI{4}{K}$ stage, and by 10 dB at the 1K stage.
The data qubits are equipped with local charge lines for independent single-qubit XY gates.
The specific control and measurement equipment used throughout the experiment is summarized in Table~\ref{tab:equipment}.
The relevant parameters of the device used in the experiment are summarized in Table~\ref{tab:params}.

\begin{table}[h!]
\centering
\begin{tabular}{p{4cm} p{3.5cm} p{2cm}}
    \hline
    \hline
    Component       & Manufacturer     & Model   \\
    \hline
    Dilution Refrigerator & Bluefors   & XLD1000 \\
    RF Source       & Rohde \& Schwarz & SGS100A  \\
    DC Source       & QDevil           & QDAC    \\
    Control Chassis & Keysight         & M9019A  \\
    AWG             & Keysight         & M3202A  \\
    ADC             & Keysight         & M3102A  \\
    \hline
    \hline
\end{tabular}
\caption{\textbf{Summary of control equipment.} The manufacturers and model numbers of experimental control equipment.}
\label{tab:equipment}
\end{table}

\begin{table}[h!]
    \centering
\begin{tabular}{p{3.5cm} p{1.3cm} p{1.3cm} p{1.3cm} p{1.3cm} p{1.3cm} p{1.3cm} p{1.3cm} p{1.3cm}}
    \hline
    \hline
    Parameter & $\textrm{Q}_1$ & $\textrm{Q}_2$ & $\textrm{Q}_3$ & $\textrm{Q}_4$ & $\textrm{Q}_5$ & $\textrm{Q}_6$ & $\textrm{Q}_7$ & $\textrm{Q}_8$\\
    \hline
    Frequency ($\SI{}{\GHz}$) & 5.0 & 5.0 & 4.9 & 4.67 & 5.0 & 5.0 & 4.9 & 4.67\\ 
    $\gamma/2\pi$ ($\SI{}{\MHz}$) & 17.7 & 17.3 & - & - & 17.9 & 17.1 & - & -\\
    $T_1$ ($\SI{}{\us}$) & - & - & 7.9 & 4.4 & - & - & 8.1 & 6.0\\
    $T_2^*$ ($\SI{}{\us}$) & - & - & 3.2 & 2.0 & - & - & 4.5 & 5.4\\
    \hline
    \hline
\end{tabular}
    \caption{\textbf{Summary of device parameters.} The operational qubit frequencies, qubit-waveguide coupling strengths $\gamma$, and $T_1$ and $T_2^*$ of the data qubits are given for the waveguide ($\textrm{Q}_{1,2,5,6}$) and data qubits ($\textrm{Q}_{3,4,7,8}$) on both modules used throughout the experiment.}
    \label{tab:params}
\end{table}
\subsection{Error Budget Analysis}
\begin{table}[h!]
    \centering
    \begin{tabular}{p{5cm} p{3cm} R{4cm} R{4cm} }
    \hline
    \hline
     Loss Mechanisms & Measurement Type &  Leftward Protocol & Rightward Protocol\\
     \hline 
     Directionality Error & Field Amplitude & 1.3\% & 1.4\%\\
     Missed Absorption & Field Amplitude & 2.9\%&5.7\%\\
     %\hline
     \textbf{Total Coherent Loss} &  & \textbf{4.2\%} & \textbf{7.1\%}\\
     \hline
     Propagation Loss & Scattering & 18\% & 18\%\\
     %Propagation Loss & Field Amplitude & 20\% & 23\%\\
     Data-Qubit Decoherence & M.E. Simulation & 6.8\% & 6.8\%\\
     %$\mathrm{Q_3}$ and $\mathrm{Q_7}$ decay & Field Measurement & 2.6\% & 1.4\%\\
     %$\mathrm{Q_3}$ and $\mathrm{Q_7}$ decay & M.E. Simulation & 2.9\% & 2.9\%\\
     Sideband Loss & Field Amplitude & 1.5\% & 1.5\%\\
     %\hline
     \textbf{Total Incoherent Loss} &  & \textbf{26.3\%} & \textbf{26.3\%}\\
     \hline
     \textbf{Total Photon Loss} &  & \textbf{30.5\%} & \textbf{33.4\%}\\
     \hline
     \hline
 \end{tabular}
    \caption{\textbf{Summary of losses in the directional emission and absorption protocol.}}
    \label{tab:losses}
\end{table}
\begin{table}[h!]
    \centering
    \begin{tabular}{p{5cm} p{3cm} R{4cm} R{4cm} }
    \hline
    \hline
     Loss Mechanisms & Measurement Type &  Leftward Protocol & Rightward Protocol\\
     \hline 
     Propagation Loss & Field Amplitude & 20\% & 23\%\\
     $\mathrm{Q_3}$ and $\mathrm{Q_7}$ decay & Field Measurement & 2.6\% & 1.4\%\\
     $\mathrm{Q_3}$ and $\mathrm{Q_7}$ decay & M.E. Simulation & 2.9\% & 2.9\%\\
     %\hline
     \hline
     \hline
 \end{tabular}
    \caption{\textbf{Alternate incoherent loss assessments of the directional emission and absorption protocol.}}
    \label{tab:losses_alt}
\end{table}
We quantify coherent protocol losses with the measured photon-field amplitudes at both ends of the waveguide following separate emission and absorption experiments.
For example, in the case of leftward emission (module B $\rightarrow$ module A), to estimate photon loss caused by imperfect directionality, we emit photons from module B in the rightward direction and calculate the integrated coherent power $P_\mathrm{R, E} = \int_W |\langle \hat a_\mathrm{R, E}(\omega) \rangle|^2 d\omega$ to calibrate the gain of the rightward detector in the interval $W \in [\omega_\mathrm{ph}-\delta \omega, \omega_\mathrm{ph}  + \delta \omega]$, where $\omega_\mathrm{ph}/2\pi = 5.0$ GHz and $\delta \omega/2\pi = 40$ MHz. 
We compare this to the integrated power of the photon field on the same detector following the absorption experiment $P_\mathrm{R, A} = \int_W |\langle \hat a_\mathrm{R, A}(\omega) \rangle|^2 d\omega$.
We calculate an upper bound for coherent photon loss due to directionality error of $P_\mathrm{R, A}/P_\mathrm{R, E} = 1.3 \%$ for the leftward protocol.
The directionality error of the rightward protocol (module A $\rightarrow$ module B) is $1.4 \%$.

Similarly, the protocol is subject to another form of coherent loss, which we call missed absorption.
For leftward emission, this error is characterized by the integrated power measured by the leftward detector after absorption, $P_\mathrm{L, A} = \int_W |\langle \hat a_\mathrm{L, A}(\omega) \rangle|^2 d\omega$.
To calculate the missed absorption loss, we emit photons in the leftward direction and detune the absorber module, placing it in transparency mode.
We measure the integrated power on the leftward detector $P_\mathrm{L, E} = \int_W |\langle \hat a_\mathrm{L, E}(\omega) \rangle|^2 d\omega$ to calculate the missed absorption loss $P_\mathrm{L, A}/P_\mathrm{L, E} = 2.9 \%$.
Likewise, the missed absorption loss resulting from the rightward protocol is $5.7 \%$.
In summary, the total coherent loss to the waveguide is $4.2\%$ in the leftward direction and $7.1\%$ in the rightward direction.

With photon-field measurements, we also estimate incoherent photon loss $\zeta_\mathrm{L/R}$ along the waveguide in between the modules.
Though we use a superconducting NbTi coaxial cable between microwave packages, the printed-circuit-board traces are made of gold-plated copper, with additional losses introduced by the ceramic dielectric.
Similarly, the K connectors on the packages are made of gold-plated beryllium copper.
To measure $\zeta_\mathrm{L}$, we emit photons from both modules individually in the leftward direction and calculate the ratios of their integrated powers measured at the leftward detector.
We measure $\zeta_\mathrm{L} = 20\%$ and $\zeta_\mathrm{R}= 18\%$, noting that we are not accounting for directionality error.
Similarly, assuming reciprocity $\zeta = \zeta_\mathrm{L}= \zeta_\mathrm{R}$, we estimate the loss between the waveguide $\zeta = 18\%$ (-0.86 dB) with single-qubit resonant scattering measurements as a function of probe power. 

Using master-equation simulations, we estimate data-qubit dephasing and lifetime contribute roughly $6.8\%$ to incoherent photon loss.
%Pure dephasing on the data qubits disrupts the interference effect critical to directional emission.
We estimate that the data-qubit lifetimes are limited by direct decay to the waveguide.
To illustrate this, in the field amplitude measurement following the absorption experiment, we observe a signal component at 4.9 GHz, which is the frequency of the data qubits (Q$_3$ and Q$_7$) closest to the photon frequency of $\omega_\mathrm{ph}/2\pi = $ 5.0 GHz.
From the integrated power at the data qubit frequency, we calculate a photon loss of $2.6\%$ ($1.4\%$) during the leftward (rightward) emission experiment.
For comparison, we find the lifetimes of these two data qubits exclusively contribute $2.9\%$ in photon loss in simulation.

In addition, the parametric pulses used to release the photon to the waveguide drive higher-order sideband transitions, which also appear in the field-amplitude signals.
The integrated power loss at the frequency of the visible sideband at 5.1 GHz is $1.5\%$ in both directions.
We estimate a total incoherent photon loss of 26.3\%, including contributions from propagation loss, data-qubit decoherence, and sideband~loss.
 
In total, we account for $30.5\%\, (33.4\%)$ photon loss in the leftward (rightward) direction. These known sources of loss are summarized in Table~\ref{tab:losses}.
We expect sideband transitions at frequencies outside our digitization bandwidth cause additional photon loss at the single-percent level.
The remaining  error may be due to modulation-induced error; the scaling of microwave-induced errors in superconducting circuits with drive amplitude, such as leakage to higher levels, is not completely understood~\cite{ding2023, Dumas2024}.
We estimate that by improving the circuit design for optimal data qubit coherence, improving the 50-$\Omega$ match of the waveguide, and mitigating photon loss, we can achieve absorption efficiencies greater than~$90\%$.
\newpage
\subsection{Single-Qubit Scattering -- Waveguide Loss Measurement}
\begin{figure*}[h!]
    \centering
    \includegraphics[width=7in]{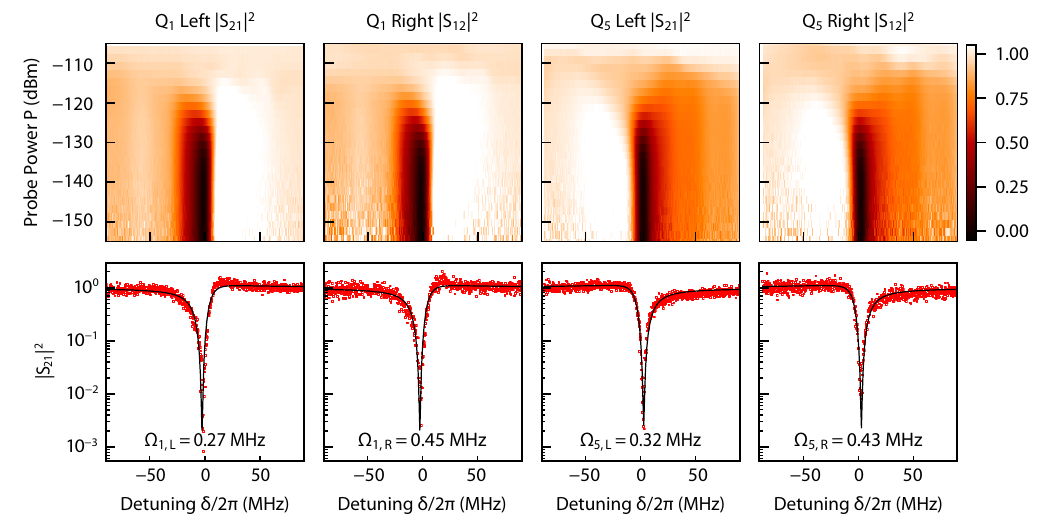}
    
    \caption{\textbf{Waveguide-qubit spectroscopy.} Magnitude of the transmission spectrum of a coherent probe incident on a qubit through the waveguide as a function of the qubit-probe detuning $\delta/2\pi$ and the probe power $P$. The measurement is repeated separately on Q$_1$ and Q$_5$ individually, which belong to different modules, with a probe propagating in the leftward or rightward direction. We fit the 2D scans to Eq.~\ref{eq:transmission} to extract the qubit-waveguide coupling rate $\gamma/2\pi \approx  17$ MHz and the drive amplitudes  $\Omega_{1,\mathrm{R/L}}$ and $\Omega_{5,\mathrm{R/L}}$ which we define at the lowest input powers in both directions. We show line cuts of the 2D scan with the Lorentzian fits at the lowest probe powers. We measure the propagation loss $\zeta \approx 18\%$ between the modules along the cable.}
    \label{fig:spec}
\end{figure*}

To fully characterize the emission and absorption protocol, we must determine the loss between the modules housed in separate microwave packaging.
One method relies on elastic scattering measurements, which enable absolute power calibration as well as the extraction of qubit-waveguide coupling and decoherence rates.

We send a coherent input tone through the waveguide which couples to a single waveguide qubit with strength $\gamma$.
For low probe powers with average photon numbers $\ll 1$, the qubit acts as a mirror to a single photon in the waveguide at a time~\cite{Hoi2011, Astafiev2010, Hoi2013}.
The master equation for driven qubit-waveguide system is ~\cite{Mirhosseini2019}
\begin{equation}
    \partial_t \hat{\rho} = -i\big[\hat{H},\hat\rho\big] + \gamma D\big[\hat{\sigma}^-\big]\hat\rho + \frac{\gamma_{\phi}}{2}D\big[\hat{\sigma}_z\big]\hat \rho.
\end{equation}
The single-qubit Hamiltonian is $\hat H = \frac{1}{2}\delta \hat\sigma_z  + \frac{1}{2}\Omega_\mathrm{p}\hat\sigma_x$, $\gamma_{\phi}$ is the pure dephasing rate, $\delta = \omega - \omega_p$ is the qubit-probe detuning, and $\Omega_\mathrm{p} = \sqrt{2\gamma P/\hbar\omega_\mathrm{p}}$ is the drive strength of the probe with power $P$.
With a rightward-propagating probe input, the output from the right end of the waveguide is
\begin{equation}
    \hat a_\mathrm{R} = \hat a_\mathrm{R}^\textrm{in} + \sqrt{\frac{\gamma}{2}}\hat \sigma^{-}.
\end{equation}
Therefore, the transmission amplitude $S_{21} = \langle\hat a_\mathrm{R}\rangle/\langle\hat a_\mathrm{R}^\textrm{in}\rangle$  is ~\cite{Mirhosseini2019}
\begin{equation}
    S_{21}(\delta, \Omega_\mathrm{p}) = 1 - \frac{\gamma(1 - i\frac{\delta}{\gamma_2})}{2\gamma_2\left(1 + \left(\frac{\delta}{\gamma_2}\right)^2 + \frac{\Omega_\mathrm{p}^2}{\gamma\gamma_2} \right)},
    \label{eq:transmission}
\end{equation}
where $\gamma_2 = \gamma/2 + \gamma_{\phi}$ is the total decoherence rate of the qubit.
Transmission measurements as a function of probe power $P$ and detuning $\delta$, as shown shown in Fig.~\ref{fig:spec} give $\gamma/2\pi  \approx 17$ MHz. Given that we use highly asymmetric junctions through a 2-$J_\mathrm{c}$ fabrication process, the pure dephasing rates $\gamma_{\phi}$ are negligible. 
We also extract the absolute power of microwave tones incident on the waveguide qubits.

To characterize the attenuation $\eta^2$ between the modules---the probabilty a photon travels from one module to another---we compare the power calibration for two corresponding waveguide qubits Q$_\mathrm{1}$ and Q$_\mathrm{5}$ in different packages.
We measure transmission as a function of probe power $P$ for each qubit in both propagation directions for a total of 4 scans to extract $\Omega_{1,\mathrm{R/L}}$, $\Omega_{5,\mathrm{R/L}}$, $P_{1,\mathrm{R/L}}$, and  $P_{5, \mathrm{R/L}}$.
Assuming $P_\mathrm{1,R}$ = $\eta^2 P_\mathrm{5,R}$, and $P_\mathrm{1,L}$ = $\eta^2  P_\mathrm{5,L}$,
\begin{equation}
    \frac{\Omega_\mathrm{1, R}^2}{\Omega_\mathrm{5, R}^2} = \frac{\gamma_1 P_\mathrm{1,R} \omega_\mathrm{5}} {\gamma_5 P_\mathrm{5,R } \omega_\mathrm{1}} 
    = \frac{\gamma_1\omega_\mathrm{5}} {\eta^2  \gamma_5\omega_\mathrm{1}} ,
\end{equation}
\begin{equation}
    \frac{\Omega_\mathrm{5, L}^2}{\Omega_\mathrm{1, L}^2} = \frac{\gamma_5 P_\mathrm{5,L} \omega_\mathrm{1}} {\gamma_1 P_\mathrm{1,L } \omega_\mathrm{5}} 
    = \frac{\gamma_5\omega_\mathrm{1}} {\eta^2  \gamma_1\omega_\mathrm{5}} .
\end{equation}
Combining these expressions, we find an analytical expression for attenuation between qubits that depends only upon the drive strength of the probe from four independent measurements,
\begin{equation}
    \eta^2  = \frac{\Omega_\mathrm{1, L}\Omega_\mathrm{5, R}}{\Omega_\mathrm{1, R}\Omega_\mathrm{5, L}}.
\end{equation}
In our system, the classical propagation and scattering loss between the qubits on different modules is $\zeta = 1-\eta^2  = 18\%$.
This is the most significant contribution to overall photon loss during the emission and absorption protocol. 
\newpage
\subsection{System Model}
\subsubsection{Chiral Waveguide Quantum Electrodynamics (wQED)}
To formulate a master equation model for wQED, we assume that the waveguide constitutes a Markovian environment---i.e. the waveguide hosts a 1D photonic continuum of time-independent modes. This approximation holds in the regime where the bandwidth of the continuum is much larger than the bandwidth of the emitted photons, which allows us to assume instantaneous interactions between atoms and the continuum of modes. This is valid for our system, since the inter-module separation is much smaller than the emitted photon bandwidth ($d\approx10 \;\text{cm}<<2\pi\nu/\gamma_\mathrm{ph}\approx17 \;\text{m}$, where $\nu$ is the speed of light in the waveguide and $\gamma_\mathrm{ph}/2\pi = 7$ MHz is the photon linewidth).  The photon travel time between modules ($\Delta t=d/\nu= 0.84 \; \text{ns}$) is negligible compared to the temporal extent of the photon wavepacket ($2\pi/\gamma_\mathrm{ph}= 90\; \text{ns}$). Therefore, the waveguide qubits on different modules interact through the waveguide modes simultaneously.

By taking the Markov approximation, the master equation for a general system of resonant waveguide qubits, each coupled bidirectionally to a common waveguide at a total rate $\gamma$, can be written in the waveguide-qubit frame as \cite{Pichler2015}
\begin{equation}
    \dot{\hat{\rho}} = -i\left[\frac{\gamma}{4}\sum_{j,l}\sin(k|x_j-x_l|)(\hat{\sigma}^\dag_j\hat{\sigma}_l + \hat{\sigma}^\dag_j\hat{\sigma}_l), 
 \hat{\rho}\right] + \gamma\sum_{j,l}\cos(k|x_j-x_l|)\mathcal{D}[\hat{\sigma}_j, \hat{\sigma}_l]\hat{\rho}.
\label{bidir_me}
\end{equation}
where $x$ indicates the position of the atom along the waveguide. The sinusoidal Hamiltonian term governs the waveguide-mediated coherent exchange interactions while the cosinusoidal Lindbladian term (when $j \ne l$) describes the correlated decay effects. 

Alternatively, the chiral wQED formalism considers the left- and right-propagating modes in the waveguide independently.
The chiral master equation is written as
\begin{equation}
    \dot{\hat{\rho}} = -\frac{i}{\hbar}[\hat{H}_\mathrm{L} + \hat{H}_\mathrm{R}, \hat{\rho}] + \gamma_\mathrm{L}\mathcal{D}[\hat{c}_\mathrm{L}]\hat{\rho} + \gamma_\mathrm{R}\mathcal{D}[\hat{c}_\mathrm{R}]\hat{\rho},
\label{chiral_me}
\end{equation}
where $\gamma_\mathrm{L}$ ($\gamma_\mathrm{R}$) describes the rate of emission of the atoms into the leftwards (rightwards) propagating modes of the waveguide.
$\hat{c}_\mathrm{L} = \sum_je^{ikx_j}\hat{\sigma}_j$ and $\hat{c}_\mathrm{R} = \sum_je^{-ikx_j}\hat{\sigma}_j$ are the leftwards and rightwards collective jump operators for the system.
The coherent waveguide-mediated interactions are described by the terms 
\begin{equation}
    \hat{H}_\mathrm{L} = -\frac{i\hbar\gamma_\mathrm{L}}{2}\sum_{j<l}(e^{ik|x_l-x_j|}\hat{\sigma}_j^\dag\hat{\sigma}_l - \text{h.c.})
\end{equation}
\begin{equation}
    \hat{H}_\mathrm{R} = -\frac{i\hbar\gamma_\mathrm{R}}{2}\sum_{j<l}(e^{ik|x_l-x_j|}\hat{\sigma}_l^\dag\hat{\sigma}_j - \text{h.c.}).
\end{equation}
In the bidirectional limit, where $\gamma_\mathrm{L} =\gamma_\mathrm{R}$, the leftwards and rightwards phase factors interact to give rise to the sinusoidal distance-dependent waveguide-mediated interaction and correlated decay terms, reducing the chiral master equation reduces to \ref{bidir_me}.
In the opposite limit, where each atom is perfectly chiral, either $\gamma_\mathrm{L}$ or $\gamma_\mathrm{R}$ is equal to zero. When $\gamma_\mathrm{R} = 0$, the master equation reduces to \begin{equation}
    \dot{\hat{\rho}} = -\frac{i}{\hbar}[\hat{H}_\mathrm{L}, \hat{\rho}] + \gamma_\mathrm{L}\mathcal{D}[\hat{c}_\mathrm{L}]\hat{\rho}.
\end{equation}
In this regime, the interactions of an atom with the waveguide modes and the interactions between atoms are unidirectional; this system is often referred to as \textit{cascaded.}
Since only the leftwards phase factor ($e^{+ikx}$) remains, there are no distance-dependent dynamics and the phase is trivial.
While in the bidirectional case, only the distance between the atoms, and not the ordering, matters; in the cascaded case, the ordering is critical, since the output of each atom can only drive other atoms downstream without back-action.
This non-Hermitian effect  can only be obtained from an open system. 

In our system, each qubit is coupled bidirectionally to the waveguide.
However, we engineer the interactions within a pair of waveguide qubits so that each module behaves as a perfectly chiral atom. 

\subsubsection{Chiral Scattering}
To better understand how our modules function as perfectly chiral atoms, we can investigate the scattering response of the two-waveguide-qubit system.
This system is made up of two resonant qubits coupled bidirectionally to a common waveguide at a total rate $\gamma$ ($\gamma_\mathrm{R} = \gamma_\mathrm{L} = \gamma / 2$).
The waveguide qubits are positioned a distance of $\lambda/4$ apart, where $\lambda$ is the wavelength corresponding to the qubits' frequency.
The propagation phase between the two waveguide qubits is $\phi = k\Delta x = \frac{\pi}{2}$.
At this distance, the correlated decay term is equal to zero and the waveguide-mediated interaction is equal to $\gamma / 2$, which we cancel with a tunable coupler. 

We define leftwards and rightwards collective decay operators for this system, $\hat{c}_\mathrm{L} = \hat{\sigma}_1^- + e^{i\frac{\pi}{2}}\hat{\sigma}_2^-$ and $\hat{c}_\mathrm{R} = \hat{\sigma}_1^- + e^{-i\frac{\pi}{2}}\hat{\sigma}_2^-$, and  write the leftwards-driven system Hamiltonian,
\begin{align}
    \hat{H} &= -\frac{\delta}{2}(\hat{\sigma}_z^1 + \hat{\sigma}_z^2) + ia_{\mathrm{in}}\sqrt{\frac{\gamma}{2}}\big{[}(\hat{\sigma}_1^- - \hat{\sigma}_1^+) + (e^{i\frac{\pi}{2}}\hat{\sigma}_2^- - e^{-i\frac{\pi}{2}}\hat{\sigma}_2^+)\big{]} \\
    &= -\frac{\delta}{2}(\hat{\sigma}_z^1 + \hat{\sigma}_z^2) + ia_{\mathrm{in}}\sqrt{\frac{\gamma}{2}}(\hat{c}_\mathrm{L} - \hat{c}_\mathrm{L}^\dag)\,,
\end{align}
and the input-output equations as
\begin{equation} \
\begin{split}
    &\hat{a}^\mathrm{out}_\mathrm{L} = \hat{a}_\mathrm{L}^\mathrm{in} + \sqrt{\frac{\gamma}{2}}\hat{c}_\mathrm{L}\,,\\
    &\hat{a}^\mathrm{out}_\mathrm{R} = \sqrt{\frac{\gamma}{2}}\hat{c}_\mathrm{R}\,,
    \label{a_R_cgm}
\end{split}
\end{equation}
where $\delta =  \omega_q - \omega_p$ and $\langle \hat a_\mathrm{L}^\mathrm{in} \rangle = i\sqrt{\frac{P}{\hbar\omega}}$ is the drive amplitude, which must be in the low power regime with $\langle \hat a_\mathrm{L}^\mathrm{in} \rangle \ll 1$ to avoid saturating the qubit.
Using the Heisenberg equations of motion $\dot{\hat{\sigma}}^- _j = -i[\hat{\sigma}^-_j, \hat{H}] - (\gamma + \Gamma_\mathrm{nr})\hat{\sigma}^-_j$, where $\Gamma_\mathrm{nr}$ is the non-radiative decay rate, we take the steady state approximation $\dot{\hat{\sigma}}^- _j = 0$ to find
\begin{equation}
    \langle\hat{\sigma}^-_1\rangle = \frac{-\langle \hat a_\mathrm{L}^\mathrm{in} \rangle\sqrt{\frac{\gamma}{2}}}{i\delta+(\gamma + \Gamma_\mathrm{nr})/2}\,,
\end{equation}
\begin{equation}
        \langle\hat{\sigma}^-_2\rangle = \frac{i\langle \hat a_\mathrm{L}^\mathrm{in} \rangle\sqrt{\frac{\gamma}{2}}}{i\delta+(\gamma + \Gamma_\mathrm{nr})/2}\,.
\end{equation}
Plugging these results into \ref{a_R_cgm}, we obtain
\begin{equation}
    t = \frac{\langle \hat{a}^\mathrm{out}_\mathrm{L} \rangle}{\langle \hat{a}_\mathrm{L}^\mathrm{in}\rangle} = 1 - \frac{\gamma}{i\delta+(\gamma + \Gamma_\mathrm{nr})/2}\,,
\end{equation}
\begin{equation}
        r = \frac{\langle \hat{a}^\mathrm{out}_\mathrm{R} \rangle}{\langle \hat{a}_\mathrm{L}^\mathrm{in}\rangle} = 0\,.
\end{equation}
Comparing these results to the scattering response of a leftwards driven general chiral atom \cite{Lodahl2017},
\begin{equation}
    t = \frac{\langle \hat{a}^\mathrm{out}_\mathrm{L} \rangle}{\langle \hat{a}^\mathrm{in}_\mathrm{L}\rangle}  = 1 - \frac{\gamma_\mathrm{L}}{i\delta+\Gamma_{tot}/2}\,,
\end{equation}
\begin{equation}
        r = \frac{\langle \hat{a}^\mathrm{out}_\mathrm{R} \rangle}{\langle \hat{a}^\mathrm{in}_\mathrm{L}\rangle} = \frac{-\sqrt{\gamma_\mathrm{R}\gamma_\mathrm{L}}}{i\delta+\Gamma_\mathrm{tot}/2}\,,
\end{equation}
we see that our system has a scattering response identical to that of a perfectly cascaded atom---the effective $\gamma_\mathrm{L}$ of the module is equal to $\gamma$, while the effective $\gamma_\mathrm{R}$ is equal to zero.
The cascaded scattering response for the lossless system is characterized by unity transmission with a phase change of $2\pi$ and zero reflection.
If we were to drive the system to the right instead, we would find the opposite response;  the effective $\gamma_\mathrm{L}$ would be zero, while the effective $\gamma_\mathrm{R}$ would be equal to $\gamma$. 

Furthermore, note that the rightwards emitting state, $|\psi_\mathrm{R}\rangle = \frac{1}{\sqrt{2}}(|10\rangle + e^{i\frac{\pi}{2}} |01\rangle)$,  is in the null space of the leftwards collective decay operator, $\hat{c}_\mathrm{L}|\psi_\mathrm{R}\rangle = 0$.
Therefore, this state can be considered to be subradiant in the leftwards direction, since destructive interference inhibits decay into the ground state via the waveguide.
Similarly, the leftwards emitting state, $|\psi_\mathrm{L}\rangle = \frac{1}{\sqrt{2}}(|10\rangle + e^{-i\frac{\pi}{2}} |01\rangle)$, is in the null space of the rightwards collective decay operator, $\hat{c}_\mathrm{R}|\psi_\mathrm{L}\rangle = 0$.  

We can understand the two-waveguide-qubit system as \textit{selectively perfectly chiral}: if we drive to the left (right) or prepare the system in the state $|\psi_\mathrm{L}\rangle$ ($|\psi_\mathrm{R}\rangle$), the system behaves as a single atom with perfect chiral coupling to the leftward (rightward) modes in the waveguide.
However, this behavior is only valid in the single-photon regime, unless we engineer a large cross-Kerr interaction between the two waveguide qubits in order to enforce a single-photon limit on the module.

\subsubsection{Cascaded Emission and Absorption Model}
To simulate emission and absorption, we construct a master equation model for the entire eight qubit system.
We first consider the four waveguide qubits. Assuming that all qubits are resonant, the master equation of this sub-system in the qubit frame is 
\begin{multline}
    \dot{\hat{\rho}} = -\frac{i}{\hbar}\Bigl[\hat{H}_\mathrm{J}, \hat{\rho}\Bigl] +\gamma\Biggl[\;\sum_{j\in\{1, 2, 5, 6\}}\mathcal{D}[\hat{\sigma}_j^-]\hat{\rho}  + \cos(kd)(\mathcal{D}[\hat{\sigma}_1^-,\hat{\sigma}_5^-]\hat{\rho} + \mathcal{D}[\hat{\sigma}_2^-, \hat{\sigma}_6^-]\hat{\rho} + \text{h.c.} ) \\
   + \cos(kd + \frac{\pi}{2})(\mathcal{D}[\hat{\sigma}_1^-, \hat{\sigma}_6^-]\hat{\rho} + \text{h.c.}) + \cos(kd - \frac{\pi}{2})(\mathcal{D}[\hat{\sigma}_2^-, \hat{\sigma}_5^-]\hat{\rho} + \text{h.c.}) \Biggl].
\label{bidir_ME}
\end{multline}
The waveguide-mediated interaction term $\hat{H}_\mathrm{J}$ is
\begin{multline}
\hat{H}_\mathrm{J} = \frac{\gamma}{2}\Bigl(
    \sin(kd)(\hat{\sigma}_1^+\hat{\sigma}_5^- + \hat{\sigma}_2^+ \hat{\sigma}_6^- + \text{h.c}.) + \sin\bigl(kd + \frac{\pi}{2}\bigl)(\hat{\sigma}_1^+\hat{\sigma}_6^- + \text{h.c}.) 
    +\sin\bigl(kd - \frac{\pi}{2}\bigl)(\hat{\sigma}_2^+\hat{\sigma}_5^- + \text{h.c}.)\Bigl).
\end{multline}
Waveguide-mediated interactions and correlated decay only occur between qubits on separate modules, since we cancel intra-module qubit coupling and there is zero correlated decay between qubits spaced $\lambda/4$ apart on the same module.
To better understand how the four-waveguide-qubit system leads to cascaded interactions, we can instead treat each module as a unit, defining leftwards and rightwards collective jump operators for modules A and B
\begin{equation}
\begin{split}
    &\hat{c}_\mathrm{L,A}^- = \hat{\sigma}_1^-+e^{i\frac{\pi}{2}}\hat{\sigma}_2^-\;\;\;\;\;\;\;\;\hat{c}_\mathrm{R,A}^- = \hat{\sigma}_1^-+e^{-i\frac{\pi}{2}}\hat{\sigma}_2^-\,, \\
    &\hat{c}_\mathrm{L,B}^- = \hat{\sigma}_5^-+e^{i\frac{\pi}{2}}\hat{\sigma}_6^-\;\;\;\;\;\;\;\;\hat{c}_\mathrm{R,B}^- = \hat{\sigma}_5^-+e^{-i\frac{\pi}{2}}\hat{\sigma}_6^-\,.
\end{split}
\label{jump ops}
\end{equation}
This allows us to  rewrite \ref{bidir_ME} in a manner that mimics the chiral wQED formalism 
\begin{equation}
    \dot{\hat{\rho}} = -\frac{i}{\hbar}[\hat{H}_\mathrm{L} + \hat{H}_\mathrm{R}, \hat{\rho}] + \frac{\gamma}{2}\Bigl(\mathcal{D}[\hat{c}_\mathrm{L,A}^- + e^{ikd}\hat{c}_\mathrm{L,B}^-]\hat{\rho} + \mathcal{D}[\hat{c}_\mathrm{R,A}^- + e^{-ikd}\hat{c}_\mathrm{R,B}^-]\hat{\rho}\Bigl)\,,
\label{2CGM_chiral_me}
\end{equation}
where
\begin{equation}
\begin{split}
    &\hat{H}_\mathrm{L} = \frac{-i\gamma}{4}\big{(}e^{ikd}\hat{c}_\mathrm{L,A}^+\hat{c}_\mathrm{L,B}^- - e^{-ikd}\hat{c}_\mathrm{L,A}^-\hat{c}_\mathrm{L,B}^+\big{)} \\
    & \hat{H}_\mathrm{R} = \frac{-i\gamma}{4}\big{(}e^{ikd}\hat{c}_\mathrm{R,A}^-\hat{c}_\mathrm{R,B}^+ - e^{-ikd}\hat{c}_\mathrm{R,A}^+\hat{c}_\mathrm{R,B}^-\big{)}\,.
\end{split}
\label{HL HR}
\end{equation}
Because $\hat{c}_\mathrm{L,A}|\psi_\mathrm{R,A}\rangle = 0$ , if we prepare module A in the rightwards-emitting state, the leftwards contributions to the system dynamics are nullified, and the system behaves in a rightwards-cascaded manner.
A photon is emitted from module A and absorbed by module B.
In this cascaded process, the distance between the two modules is irrelevant to the dynamics and only manifests as a phase in the absorbed state.
The reverse process occurs if we instead prepare module B in the leftwards-emitting state.

To account for propagation loss between the modules, we follow the approach in \cite{irfan2024, reuer2021realization} by introducing a fictitious beamsplitter between the modules that allows a photon to propagate in either direction with probability $\eta^2$.
We define SLH triplets~\cite{Kockum2018,combes}
\begin{gather}
    G_\mathrm{R(L),A(B)} = \Biggl{(}I, \sqrt{\frac{\gamma}{2}}\begin{pmatrix}\hat{c}_\mathrm{R(L),A(B)}^-\\0\end{pmatrix}, 0\Bigl{)}\,, \\
    G_\mathrm{BS} = \Biggl{(}\begin{pmatrix}
       \eta & -\sqrt{1-\eta^2} \\
       \sqrt{1-\eta^2} & \eta
   \end{pmatrix}, 0, 0\Biggl{)}\,, \\
    G_\mathrm{\phi_{WG}} = \Biggl{(}e^{ikd}, 0, 0\Biggl{)}\,,
\end{gather}
where $G_\mathrm{R(L),A(B)}$ is the rightwards (leftwards) triplet for module A (B),  $G_\mathrm{BS}$ is the beamsplitter triplet, and  $G_\mathrm{\phi_{WG}}$ is the triplet describing the acquired propagation phase between the modules.
We for the solve the rightwards cascaded system using the series product \cite{Kockum2018}
\begin{multline}
G_\mathrm{1} = G_\mathrm{R,B} \triangleleft G_\mathrm{\phi_{WG}} \triangleleft G_\mathrm{BS} \triangleleft G_\mathrm{R,A} =  \\
\Biggl{(}e^{ikd}\begin{pmatrix}
       \eta & -\sqrt{1-\eta^2} \\
       \sqrt{1-\eta^2} & \eta
   \end{pmatrix}, \begin{pmatrix}
       \sqrt{\frac{\gamma}{2}}(\eta \hat{c}_\mathrm{R,A}^- + e^{-ikd}\hat{c}_\mathrm{R,B}^- )\\  \sqrt{\frac{\gamma}{2}} \sqrt{1 - \eta^2}\hat{c}_\mathrm{R,A}^-
   \end{pmatrix}, \frac{-i\gamma\eta}{4}\big{(}e^{ikd}\hat{c}_\mathrm{R,A}^-\hat{c}_\mathrm{R,B}^+ - e^{-ikd}\hat{c}_\mathrm{R,A}^+\hat{c}_\mathrm{R,B}^-\big{)}\Biggl{)}.
\end{multline}
Similarly, the leftwards cascaded system gives 
\begin{multline}
G_\mathrm{2} = G_\mathrm{L,A} \triangleleft G_\mathrm{\phi_{WG}} \triangleleft G_\mathrm{BS} \triangleleft G_\mathrm{L,B} =  \\
\Biggl{(}e^{ikd}\begin{pmatrix}
       \eta & -\sqrt{1-\eta^2} \\
       \sqrt{1-\eta^2} & \eta
   \end{pmatrix}, \begin{pmatrix}
       \sqrt{\frac{\gamma}{2}}(\hat{c}_\mathrm{L,A}^- +\eta e^{ikd}\hat{c}_\mathrm{L,B}^- )\\  \sqrt{\frac{\gamma}{2}} \sqrt{1 - \eta^2}\hat{c}_\mathrm{L,B}^-
   \end{pmatrix}, \frac{-i\gamma\eta}{4}\big{(}e^{ikd}\hat{c}_\mathrm{L,A}^+\hat{c}_\mathrm{L,B}^- - e^{-ikd}\hat{c}_\mathrm{L,A}^-\hat{c}_\mathrm{L,B}^+\big{)}\Biggl{)}.
\label{left_SLH}
\end{multline}
Combining these two systems using the concatenation product, $G_\mathrm{1} \boxplus G_\mathrm{2}$ \cite{Kockum2018}, we find that waveguide mediated interaction Hamiltonian now has a multiplicative factor of $\eta$ , $\hat{H} = \eta(\hat{H}_\mathrm{L} + \hat{H}_\mathrm{R})$, and the Lindbladian terms are modified as such:
\begin{equation}
\begin{split}
&\hat{c}_1 = \eta\hat{c}_\mathrm{R,A}^- + e^{-ikd}\hat{c}_\mathrm{R,B}^- \\
&\hat{c}_2 = \sqrt{1-\eta^2}\hat{c}_\mathrm{R,A}^- \\
&\hat{c}_3 = \hat{c}_\mathrm{L,A}^- +\eta e^{ikd}\hat{c}_\mathrm{L,B}^-\\
&\hat{c}_4 = \sqrt{1-\eta^2}\hat{c}_\mathrm{L,B}^-\,. \\
\end{split}
\label{lindblads}
\end{equation}
$\hat{c}_1$ and $\hat{c}_2$  describe photon transfer from modules A to B, while $\hat{c}_3$ and $\hat{c}_4$  describe photon transfer from modules B to A.
Therefore, a non-zero $\eta$ generates photon loss by creating an asymmetry in the collective loss operators $\hat{c}_1$ and $\hat{c}_3$ and by introducing operators $\hat{c}_2$ and $\hat{c}_4$, which describe on-site loss on the upstream qubits. 

We incorporate the data-waveguide qubit interactions by adding Hamiltonian terms describing the parametric coupling between each data-waveguide qubit pair.
We modulate the tunable coupler at the data-waveguide qubit detuning with shaped fast-flux pulses to generate a time-dependent coupling $g(t)$ to emit a symmetric photon.
We use the reverse pulses $g(-t)$ to absorb the photon.
In the detuning frame and under the rotating wave approximation, these terms are expressed as
\begin{multline}
        H = g_1(t)(\sigma_1^+\sigma_3^- + \sigma_3^+\sigma_1^-) + g_2(t)(\sigma_2^+\sigma_4^- + \sigma_4^+\sigma_2^-) + g_3(-t)(\sigma_5^+\sigma_7^- + \sigma_7^+\sigma_5^-) + g_4(-t)(\sigma_6^+\sigma_8^- + \sigma_8^+\sigma_6^-).
\end{multline}
For simplicity, the initial physics-based experimental calibration starts with pulse envelopes that generate a symmetric, hyperbolic secant photon envelope $f(t)$ with bandwidth $\gamma_\mathrm{ph}$,
\begin{equation}
    f(t) = \frac{\sqrt{\gamma_\mathrm{ph}}}{2}\text{sech}\Bigl{(}\frac{\gamma_\mathrm{ph} t}{2}\Bigl{)}.
\end{equation}
To create this photon wavepacket, we must apply coupling pulses with the envelope $g(t)$ 
\begin{equation}
    g(t)=\frac{\gamma_\mathrm{ph}}{4\text{cosh}\Bigl{(}\frac{\gamma_\mathrm{ph}t}{2}\Bigl{)}
}\frac{(1+e^{\gamma_\mathrm{ph}t})\gamma/\gamma_\mathrm{ph}+1-e^{\gamma_\mathrm{ph}}t}{\sqrt{(1+e^{\gamma_\mathrm{ph}t})\gamma/\gamma_\mathrm{ph}-e^{\gamma_\mathrm{ph}t}}}\,,
\end{equation}
where $\gamma$ is the waveguide-qubit coupling rate.
The speed of the emission protocol is limited by $\gamma$; $\gamma_\mathrm{ph}\leq\gamma$.
When $\gamma_\mathrm{ph} = \gamma$, $g(t)$ reduces to a hyperbolic secant function. 

Finally, in order to add the effects of data qubit non-radiative decay and pure dephasing to the model, we add the following Lindbladian terms for each data qubit to the master equation
\begin{equation}
    \sum_{i\in\{3, 4, 7, 8\}} \Gamma_{\mathrm{nr}, i}\mathcal{D}[\sigma_i^-]\hat{\rho} + \frac{\gamma_{\phi, i}}{2}\mathcal{D}[\sigma_z^i]\hat{\rho}.
\end{equation}
By initiating the data qubits in module A (B) in $|\psi_\mathrm{R,A}\rangle$  ($|\psi_\mathrm{L,B}\rangle$) and using appropriately shaped coupling pulses $g(t)$, we observe single-photon transfer between the modules in simulation.
We use the full eight-qubit master equation model to fit to the results in Fig. \ref{fig:fig2} and extract the contribution of decoherence to photon loss.
\newpage
\subsection{Two-module Scattering Model}
\begin{figure}[h!]
     \centering
     \includegraphics[width = 7in]{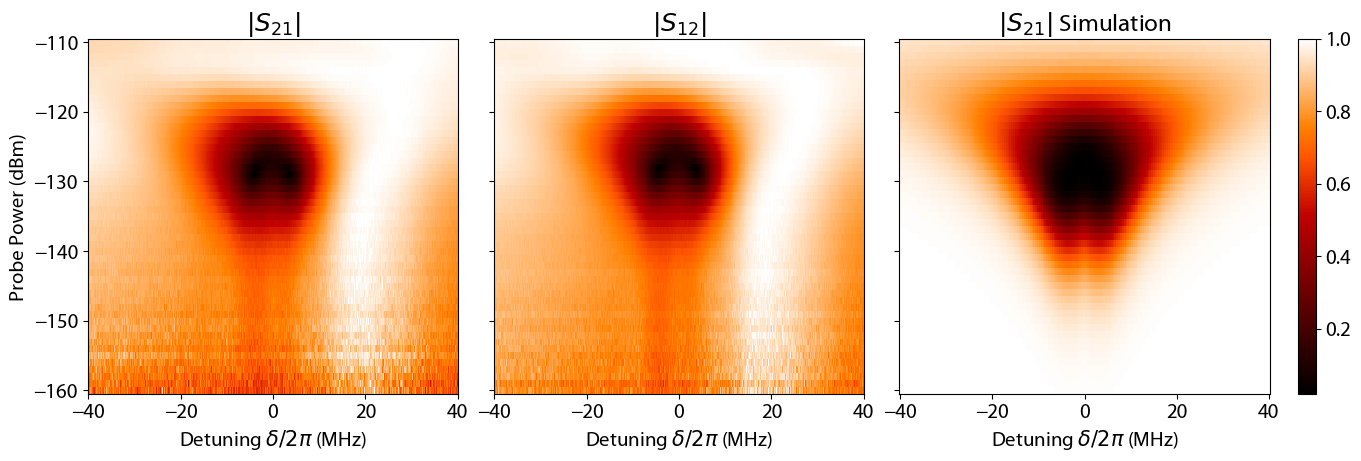}
     \caption{\textbf{Scattering response of four-waveguide-qubit system.} Transmission spectrum of a coherent probe through the waveguide incident on the four waveguide qubits across both modules,  as a function of the qubit-probe detuning $\delta/2\pi$ and the probe power $P$,  in both propagation directions (left, center) and in simulation (right). Waveguide-mediated interactions between waveguide qubits on the same module are cancelled using tunable couplers. Each module functions as a cascaded atom, resulting in approximately unity transmission. At higher powers, the cascaded atom approximation breaks as waveguide-qubit pairs are doubly populated, resulting in transmission dips. In this regime the spectrum exhibits a subtle distance-dependence, which allows us to estimate that our inter-module distance is near a multiple of $\lambda/8$ by matching to simulation. At very high powers, all waveguide qubits are saturated, and the spectrum returns to unity transmission. }
     \label{fig:four_qubit_scattering}
 \end{figure}
We also use a master equation model to simulate the four-emitter system when driven through the waveguide.
We can derive the master equation using an SLH approach similar to that of the previous section.
The module triplets now include an extra Hamiltonian term in the probe frame
\begin{equation}
    G_\mathrm{R(L),A(B)} = \Biggl{(}I, \sqrt{\frac{\gamma}{2}}\begin{pmatrix}\hat{c}_\mathrm{R(L),A(B)}^-\\0\end{pmatrix}, \frac{\delta}{2}\big{(}\hat{\sigma}_{1(5)}^z  +\hat{\sigma}_{2(6)}^z\big{)}\Biggl{)}\,,
\end{equation}
where $\delta$ is the emitter-drive detuning. The drive triplet is defined as 
\begin{equation}
    G_\mathrm{drive} = \Big{(}I, \begin{pmatrix}
    \alpha \\ 0
    \end{pmatrix}
    , 0\Big{)},
\end{equation}
where $\alpha = \langle \hat a_\mathrm{R}^\mathrm{in}\rangle$ is the amplitude of the coherent rightward drive.
For the rightwards-driven cascaded system, we solve $G_3 = G_\mathrm{R,B} \triangleleft G_\mathrm{\phi_{WG}} \triangleleft G_\mathrm{BS} \triangleleft G_\mathrm{R,A} \triangleleft G_\mathrm{drive}$ to find
\begin{gather}
  \hat{L}_3 = \begin{pmatrix}
       \eta \alpha + \sqrt{\frac{\gamma}{2}}(\eta \hat{c}_\mathrm{R,A}^- + e^{-ikd}\hat{c}_\mathrm{R,B}^- )\\  \sqrt{\frac{\gamma}{2}} \sqrt{1 - \eta^2}\hat{c}_\mathrm{R,A}^-
   \end{pmatrix} \\
   \hat{H}_3 = \sum_{i\in\{1, 2, 5, 6\}}\frac{\delta}{2}\hat{\sigma}_z^i + \eta\hat{H}_\mathrm{R} + \frac{1}{2}(\Omega\hat{c}_\mathrm{R,A}^+ + \Omega^*\hat{c}_\mathrm{R,A}^-) + \frac{1}{2}\eta(e^{ikd}\Omega\hat{c}_\mathrm{R,B}^+ + e^{-ikd}\Omega^*\hat{c}_\mathrm{R,B}^-).
\end{gather}
To complete the system, we concatenate $G_3$ with the leftwards (un-driven) cascaded system (\ref{left_SLH}), $G_4 = G_\mathrm{3} \boxplus G_\mathrm{2}$, and calculate 
\begin{gather}
    \hat{L}_4 = \begin{pmatrix}
       \eta \alpha + \sqrt{\frac{\gamma}{2}}(\eta \hat{c}_\mathrm{R,A}^- + e^{-ikd}\hat{c}_\mathrm{R,B}^- )\\  \sqrt{\frac{\gamma}{2}} \sqrt{1 - \eta^2}\hat{c}_\mathrm{R,A}^- \\  \sqrt{\frac{\gamma}{2}}(\hat{c}_\mathrm{L,A}^- +\eta e^{ikd}\hat{c}_\mathrm{L,B}^- )\\  \sqrt{\frac{\gamma}{2}} \sqrt{1 - \eta^2}\hat{c}_\mathrm{L,B}^-
   \end{pmatrix}\\
   \hat{H}_4 = \sum_{i\in\{1, 2, 5, 6\}}\frac{\delta}{2}\hat{\sigma}_z^i + \eta(\hat{H}_\mathrm{R} + \hat{H}_\mathrm{L}) + (\Omega\hat{c}_\mathrm{R,A}^+ + \Omega^*\hat{c}_\mathrm{R,A}^-) + \eta(e^{ikd}\Omega\hat{c}_\mathrm{R,B}^+ + e^{-ikd}\Omega^*\hat{c}_\mathrm{R,B}^-)\,,
\end{gather}
where $\Omega = -i\sqrt{\frac{\gamma}{2}}\alpha$/2.
We see that the propagation loss in the interconnect also affects the drive---the Hamiltonian drive term for the downstream module B has a multiplicative factor of $\eta$, as does the drive amplitude $\alpha$ in $\hat{L}_4$.
We can simulate the system by solving the master equation with $\hat{H}_4$ and the Lindbladian terms given by $\hat{L}_4$, excluding the drive term (equivalent to \ref{lindblads}).
The steady-state solutions for $\langle\hat{c}_\mathrm{R,A}^-\rangle$ and  $\langle\hat{c}_\mathrm{R,B}^-\rangle$ are then plugged into the input-ouput equation
\begin{equation}
    \hat{a}_\mathrm{R}^\mathrm{out} = \eta \hat a_\mathrm{R}^\mathrm{in} + \sqrt{\frac{\gamma}{2}}(\eta \hat{c}_\mathrm{R,A}^- + e^{-ikd}\hat{c}_\mathrm{R,B}^- )
\end{equation}
to solve for $S_{21} = \langle\hat a_\mathrm{R}^\mathrm{out}\rangle/\langle\hat a_\mathrm{R}^\mathrm{in}\rangle$.
The simulation for our specific system is shown in Fig. \ref{fig:four_qubit_scattering} for a range of drive powers.
In the low power regime, we observe unity transmission---which is the expected behaviour of two cascaded atoms.
We use this fact to calibrate our system such that the waveguide qubits are all resonant and spaced at a $\lambda/4$ distance with zero coupling within each module. 
At higher powers, a single module can be populated with two or more excitations, causing the cascaded model to break down and resulting in dips in the transmission spectrum. 
\newpage
\subsection{Time-domain reflectometry with single, microwave photons}
\begin{figure}[h!]
     \centering
     \includegraphics[width = \textwidth]{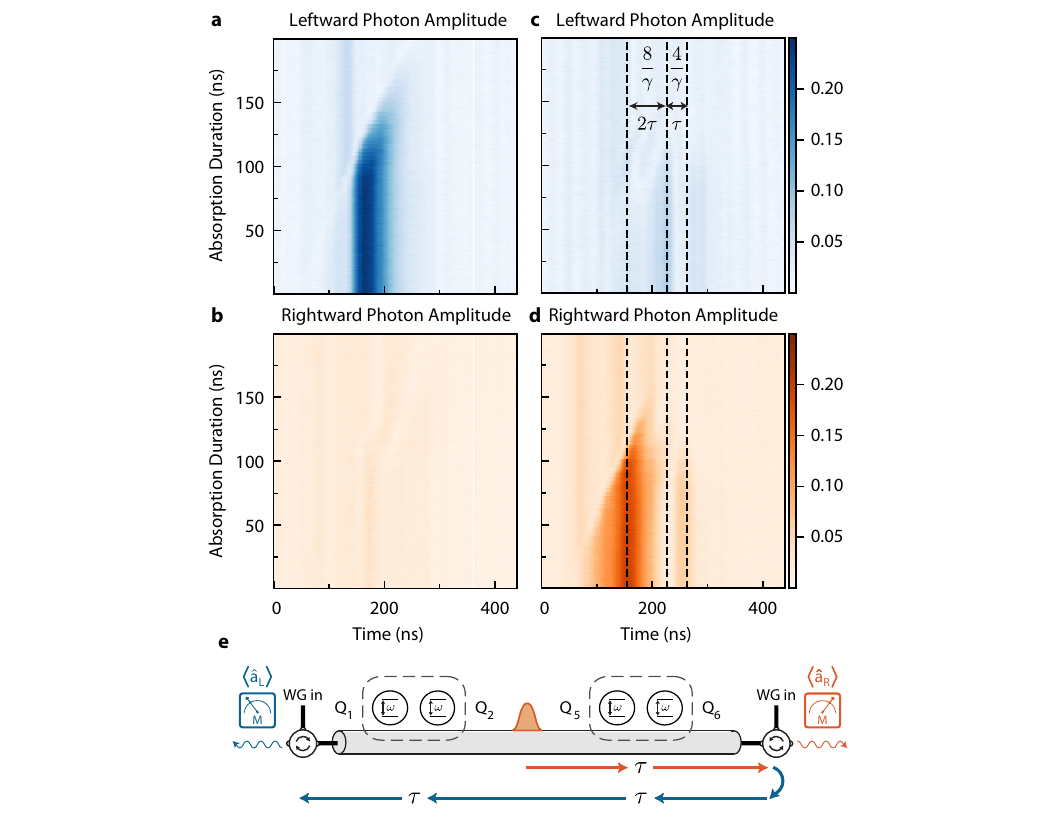}
     \caption{\textbf{Photon amplitude as a function of absorption pulse duration.}  \textbf{a,b)} Photon field amplitudes at the left and right ends of the waveguide during the leftward (\textbf{c,d} rightward) emission and absorption experiment.  We sweep the time duration of the flux pulses that absorb the photon from 0 to 200 ns. We observe the gradual absorption of the leftward photon. As indicated by the dashed lines, the unabsorbed photon sees delays that are integer multiples of $\tau = 4/\gamma$ that result from the chiral interaction with waveguide qubits on both modules~\cite{Gheeraert2020, Weiss2024}. This suggests the presence of impedance mismatches in between the modules and beyond the interconnect. When the absorption pulses reach their full duration, these signals are suppressed. \textbf{e)} Example rightward photon signal path in response to impedance mismatches downstream. Each time the photon encounters a ``transparent'' module it picks up a time delay of $\tau$. In the case of rightward emission in part \textbf{c,d}, the photon signal sees a significant impedance mismatch, the circulators for example, beyond the modules because this signal is diminished following photon absorption.}
     \label{fig:photon_amp_2D}
\end{figure}
We operate the modules in ``transparency'' mode in two different ways, either by detuning the waveguide qubits away from the directional photon frequency as done in Fig.~\ref{fig:fig2}d/e, or by keeping the waveguide qubits resonant and turning off the absorbing flux pulses as done in Fig.~\ref{fig:photon_amp_2D}.
In the latter case, the destructive interference effect guarantees unity transmission of the photon past the resonant waveguide qubits with a time delay of $4/\gamma$ and slight wavepacket distortion~\cite{Gheeraert2020,Weiss2024}.
Starting from the input-output equations of one module,
\begin{equation}
\begin{split}
        &\langle\hat{a}_\textrm{L}\rangle = \langle \hat a_\mathrm{L}^\mathrm{in}  \rangle +  \sqrt{\frac{\gamma}{2}} \left(\langle\hat{\sigma}_1^-\rangle + i\langle\hat{\sigma}_2^-\rangle \right), \\
        &\langle\hat{a}_\textrm{R}\rangle = \langle \hat a_\mathrm{R}^\mathrm{in}  \rangle +  \sqrt{\frac{\gamma}{2}} \left(\langle\hat{\sigma}_1^-\rangle -i\langle\hat{\sigma}_2^-\rangle\right),
\end{split}
\end{equation}
with the condition that an input signal with detuning $\delta$ from the qubits comes from only one direction, $\langle \hat a_\mathrm{R}^\mathrm{in}  \rangle = 0$ or  $\langle \hat a_\mathrm{L}^\mathrm{in}  \rangle = 0$, we calculate the transmission coefficient in the frequency domain,
\begin{equation}
    S_\mathrm{LR} = S_\mathrm{RL} = \frac{\langle\hat{a}_\textrm{L/R}\rangle}{ \langle \hat a_\mathrm{L/R}^\mathrm{in} \rangle} = \frac{\gamma/2   +i\delta}{\gamma/2 -i\delta}.
\end{equation}
For a small photon linewidth $\gamma_\mathrm{ph} \leq 0.5 \gamma$, which is the case in our experiment, the distortions on the wavepacket are negligible~\cite{Gheeraert2020}.
We obtain the phase imparted on the directional photon by the transparent module 
\begin{equation}
    \theta(\delta) = \arctan\big(\frac{\gamma \delta}{\gamma^2/4 - \delta^2}\big)\,,
\end{equation}
which gives the group delay of the photon wavepacket,
\begin{equation}
    \tau = \frac{\partial \theta(\delta)}{\partial \delta}\Bigg|_{\delta=0} = -\frac{4}{\gamma}.
\end{equation}
The experiments shown in Fig.~\ref{fig:photon_amp_2D} showcase several of these delays.
When the absorbing pulse duration is zero, the center of the photon wavepacket that lands at the expected detector is already $4/\gamma = 37.4$ ns delayed compared to the experiment where the waveguide qubits on the absorber module are detuned in Fig.~\ref{fig:fig2}d/e.
The first dotted line already includes a first delay imparted by the resonant waveguide qubits. 

These are not the only signal delays shown in the bottom half of these plots.
An illustrative example is shown by the rightward emission experiment in Fig.~\ref{fig:photon_amp_2D}c-e.
After the initial photon emission, there is another signal at the leftward end of the waveguide that is delayed by $8/\gamma$, which means the photon sometimes interacts with resonant modules twice more.
This implies that an impedance mismatch after the absorber module, likely a circulator, reflects the photon backwards in the leftward direction sometimes, which results in interactions with both modules on its way to the leftward waveguide detector. 
The longer absorption pulses represented by the top half of the plots suppress the reflection signals.
In other words, the module absorbs the photon before it has a chance to reflect off impedance mismatches outside the interconnect. 
The time delay imposed by the transparent module provides a means to slow down the photon pulse and observe the effect of impedance mismatches near and along the interconnect--a measurement similar to time-domain reflectometry.

\newpage
\subsection{Reinforcement Learning Optimization}
  \begin{figure}[h!]
     \centering
     \includegraphics[width = \textwidth]{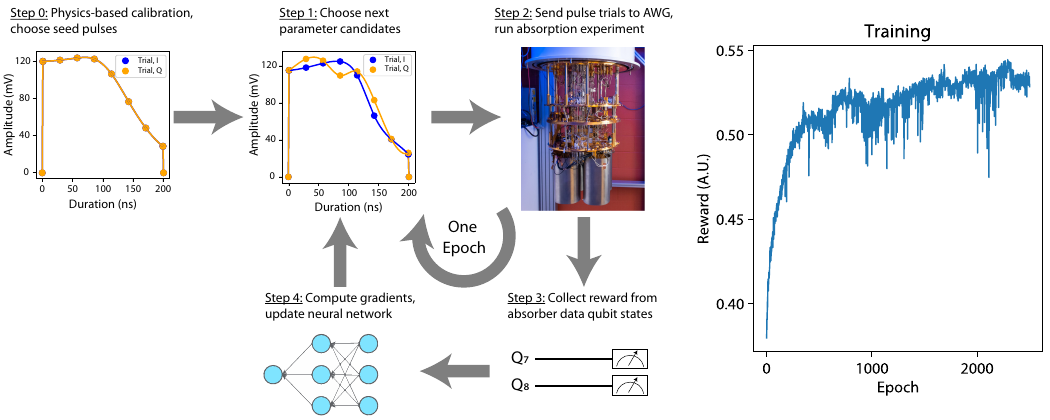}
     \caption{\textbf{Reinforcement learning optimization for maximal photon absorption.} Following a manual physics-based calibration of the shaped emission and absorption, we provide seed pulses to the reinforcement learning algorithm for all 6 protocol pulses . To allow freedom in the potential pulse shapes, we segment emission and absorption seed pulses into 8 time steps, each with an in-phase (I) and quadrature (Q) component. We optimize a total of 73 parameters, including pulse amplitudes, frequencies, and phases. We use proximal policy optimization (PPO), which then samples the next set of parameters from Gaussian distributions centered around the seed pulse parameters to form a batch of 150 trial pulse sets. We run the absorption experiment with the batch of pulse sets and collect a reward, which we define with the relation between the data qubits on the absorber module, Q$_7$ excited XOR Q$_8$ excited, after each set of trial pulses. The PPO algorithm then computes the gradients, updates neural network parameters, and then calculates a new Gaussian probability distribution for each parameter. We repeat this cycle for around 1000 epochs until the reward converges to a value, which takes around 2.5 hours.}
     \label{fig:RL}
 \end{figure}
\begin{table}[h!]
\centering
\begin{tabular}{p{5cm} p{1.3cm}}
    \hline
    \hline
    PPO Parameter& Value\\
    \hline
Learning rate & 0.005\\
Number~of~policy~updates & 20\\
Importance ratio clipping & 0.05\\
Batch size & 150\\
Number of averages & 1000\\
  Value prediction loss coefficient   & 0.5\\
Gradient clipping & 1.0\\
Log probability clipping & 0.0\\
Neural network layers & 4\\
Neural network nodes per layer & 10\\
    \hline
    \hline
\end{tabular}
\caption{
\textbf{Reinforcement learning parameters.} Hyper-parameters of the proximal policy optimization (PPO) algorithm used to optimize the emission and absorption protocol.
}\label{tab:RL}
\end{table}
We use model-free reinforcement learning to maximize absorption efficiency for both photon propagation directions.
There are six pulses involved in the protocol, each with interdependent frequencies, phases, and amplitudes.
Four of the pulses emit and absorb the photon, and the shapes of their envelopes in relation to each other is the key to maximal absorption.
We start with ideal pulse shapes that theoretically produce the photon shape $\propto \mathrm{sech}(\gamma t/2)$, which we calibrate manually.
We parameterize these four pulses with eight time steps, each with in-phase and quadrature components, and optimize a total of 73 parameters.

We use an algorithm known as proximal policy optimization (PPO), following the approach described and implemented in~\cite{sivak2022, sivak2023, ding2023}.
We adapt the code from~\cite{SivakGH}, which is built on TF-agents (Tensor Flow).
For each epoch, we send a batch of 150 trial pulses (1k averages for each trial) and run the emission and absorption experiment and simultaneously measure the populations of the data qubits on the absorber module with single-shot qubit state readout.
At the end of the protocol, the data qubits are entangled in the single-photon subspace.
We seek to maximize the degree of entanglement, so we define the reward as the sum number of counts of the two-qubit states $|01\rangle$ or $|10\rangle$.

Given the rewards, the PPO algorithm computes gradients and updates neural network parameters for the next epoch.
The neural network produces a Gaussian probability distribution for each parameter, from which the next batch of pulse trials is selected.
We run the optimization for roughly 1000 epochs and see an absorption efficiency improvement of up to 10$\%$ compared to the physics-based calibration.

The simplest manual calibration produces a symmetric photon wavepacket, such that the absorption pulses are the time-reverse of the emission pulses. Remarkably, the reinforcement learning agent produces photon shapes, as shown in Fig. 2d/e that are less symmetric than what we can calibrate by hand. This suggests that practice, a symmetric wavepacket is not necessarily optimal for maximal absorption efficiency. By distorting the control pulses and thus, the photon wavepacket, the agent apparently incorporates experimental imperfections such as control line distortions and impedance mismatches along the waveguide between the module when optimizing the wavepacket.

There could be additional distortions along the photon readout/amplification chain beyond the modules that impact the observed photon shape without impacting the emission and absorption protocol. The photon shape is important only once it arrives at the absorber module, so it could be the case that the optimizer distorts the photon to account for impedance mismatches between the modules.

Since we have control of both the absorption rate and the emission rate of the modules, we can in principle emit asymmetric wavepackets through temporally tuning the coupling strength on the emitter, and then absorb that asymmetric wavepacket by accordingly tuning the coupling strength absorption side. The optimizer takes this into account, apparently, to realize a higher efficiency.

\newpage
\subsection{Quantum Network Considerations}
\begin{table}[h!]
    \centering
\begin{tabular}{L{2.5cm} L{3cm} L{3cm} L{3cm} L{3cm} L{2.5cm}}
    \hline
    \hline
    Experimental Interconnect& Almanakly (2024)& Kurpiers (2018)~\cite{Kurpiers2018}& Zhong (2019)~\cite{Zhong2019}& Zhong (2021)~\cite{Zhong2021}& Wu (2024)~\cite{Wu2024}\\
    \hline
    \hline
    Physical Mechanism& Directional photon& Itinerant photon& Itinerant photon& Resonant mode& Router circuit\\
    \hline
    Number of  nodes per interconnect& Arbitrary number $N$ & Two& Two& Two&Single digits  \\
    \hline
    Network Connectivity& All-to-all on waveguide& All-to-all chain of circulator-node pairs& Point-to-point& Point-to-point& All-to-all for a few nodes  \\
    \hline
 Communication flow& Bidirectional& Unidirectional& 
 Bidirectional& Bidirectional&Bidirectional\\
    \hline
Distance-dependence& None& None& Yes& Yes& Yes\\
    \hline
On-chip implementation& Available& Not available& Available& Available& Available\\
    \hline
Additional notes&Enables long-distance coupling without the need for sequential iteration.& Requires lossy circulator for impedance matching. Strictly unidirectional communication.& Emission speed must be faster than travel time between nodes, constraining possible node distances and protocol speeds. &Protocol operation and speed depends on the free-spectral range of the resonator. & Only several nodes, they must be qubit-scale distance apart. Larger distances unavailable.\\
    \hline
    \hline
\end{tabular}
    \caption{\textbf{State-of-the-art superconducting quantum interconnects.} Network architecture evaluation of several select representative demonstrations of quantum interconnects.}
    \label{tab:interconnects}
\end{table}

In general, superconducting interconnects to date have not been compatible with all-to-all connectivity in both long-distance and on-chip implementations. Point-to-point communication, where an interconnect only connects two computational nodes, is not feasible for the large-scale system sizes required to realize fault-tolerant quantum computation. We compare the features of several experimental superconducting interconnects in Table~\ref{tab:interconnects}.

In addition to the advantage created by all-to-all connectivity, a larger chiral quantum network supports parallelized communication. For a given waveguide, photons can travel in two directions, doubling the channel throughput. Communication can also be multiplexed over different photon frequency bands spaced on the order of a few qubit-waveguide coupling rates (gamma/2pi = 17 MHz in this work). In a large quantum network, with the protocol described in the main text, we can simultaneously remotely entangle many pairs of modules by exploiting the time, frequency, and direction dimensions of photon propagation.

We could also generate remote, multi-partite W states over an arbitrary number of modules N by extending the remote entanglement protocol presented in the main text. By adjusting the shapes or durations of the emission pulses, we could emit a photon to the left and/or right with probability (N-1)/N. Similarly, the absorber pulses can be shaped such that each absorber module receives the photon with probability 1/N.

We also note that long-distance coupling is advantageous also for logical qubit operations of surface code logical qubits. It would enable non-nearest-neighbor coupling of logical qubits, rather than sequential nearest-neighbor operations. Additionally, many low-density parity check (LDPC) codes have encodings that require non-nearest-neighbor couplings, and these interconnects would be a means to realize such couplings efficiently. In both cases, the fact that we need not have a unique (dedicated) interconnect between every pair of physical qubits, or an interconnect limited to only a few (more than two, but fundamentally limited) physical qubits, is an advantage of our approach. It would be unwieldy to have unique interconnects between every pair of qubits we wish to connect; this clearly does not scale well.

\end{document}